\magnification=\magstep1

%--------------------------------------------------------------------------
%       various definitions to save typing
%--------------------------------------------------------------------------

\def\od{one-dimensional}

\def\pd{probability distribution}

\def\ito{It\^o}

\def\real{{\pmb{R}}}

\def\ep{\epsilon}

\def\damtp{\baselineskip=10pt
\eightsl Department of Applied Mathematics and Theoretical Physics,\break%
\eightsl University of Cambridge, Cambridge CB3 9EW, UK\par}
\def\ont{\baselineskip=10pt
\eightsl Optique nonlin\'eaire th\'eorique, \break
Universit\'e libre de Bruxelles, CP231 \break
Bruxelles 1050 BELGIUM}
\def\and{\ifnyasrefstyle \&\ \else and \fi}%
\def\regime{r\'egime}%
%
%
% to be used only in mathmode----------------------------
\def\xm{x_{\rm max}}

\def\mle{\mu |\ln \epsilon |}
\def\Or#1{{\cal O}(#1)}

\def\ha{{1 \over 2}}

%--------------------------------------------------------------------------
%       Fonts
%--------------------------------------------------------------------------
%from p.409
%\font\sixteenrm=cmr10  scaled\magstep3%
%\font\sixteenbf=cmbx10 scaled\magstep3%
%\font\sixteenit=cmti10 scaled\magstep3%
%\font\sixteensl=cmsl10 scaled\magstep3%
%\font\fourteenrm=cmr12  scaled\magstep1%
%\font\fourteenbf=cmbx12 scaled\magstep1%
%\font\fourteenit=cmti12 scaled\magstep1%
%\font\fourteensl=cmsl12 scaled\magstep1%
%\font\twelverm=cmr12%
%\font\twelvebf=cmbx12%
%\font\twelveit=cmti12%
%\font\twelvesl=cmsl12%
%\font\eightrm=cmr8%
%\font\eightbf=cmbx8%
%\font\eightit=cmti8%
%\font\eightsl=cmsl8%

% FONT DEFINITIONS
% nicked from the CUP macros

\font\sixrm=cmr6           \font\sevenrm=cmr7
\font\sixi=cmmi6           \font\seveni=cmmi7
\font\sixsy=cmsy6          \font\sevensy=cmsy7

\font\eightrm=cmr8         \font\tenrm=cmr10
\font\eighti=cmmi8         
\font\eightsy=cmsy8        
\font\eightit=cmti8        \font\tenit=cmti10
\font\eightsl=cmsl8        \font\tensl=cmsl10
\font\eightbf=cmbx8        \font\tenbf=cmbx10
\font\eighttt=cmtt8        

\font\twelverm=cmr12

%\font\twelvesy=cmsy12
\font\twelveit=cmti12
\font\twelvesl=cmsl12
\font\twelvebf=cmbx12

\font\fourteenrm=cmr12 at 14pt

%\font\fourteensy=cmsy12 at 14pt
\font\fourteenit=cmti12 at 14pt
\font\fourteensl=cmsl12 at 14pt
\font\fourteenbf=cmbx12 at 14pt

%\font\sixteenrm=cmr17 at 16pt
%\font\twentyrm=cmr17 at 20pt
%\font\sixteenrm=cmr17
%\font\twentyrm=cmr17 at 20.75pt

%\font\sixteensy=cmsy17

%\font\twentysy=cmsy17 at 20.75pt

% Poor Man's Bold -- for bold-greek or bold-italic---------------
\def\pmb#1{\setbox0=\hbox{#1}%
 \kern-.025em\copy0\kern-\wd0%
 \kern.05em\copy0\kern-\wd0%
 \kern-.025em\raise.0433em\box0}
% N.B. these big/small fonts DON'T WORK with math symbols.---------------
\def\Bigtype{%
\let\rm=\fourteenrm \let\bf=\fourteenbf%
\let\it=\fourteenit \let\sl=\fourteensl%
\baselineskip=1.44\baselineskip \rm}%
\def\bigtype{%
\let\rm=\twelverm \let\bf=\twelvebf%
\let\it=\twelveit \let\sl=\twelvesl%
\baselineskip=1.2\baselineskip \rm}%
\def\medtype{%
\let\rm=\tenrm \let\bf=\tenbf%
\let\it=\tenit \let\sl=\tensl%
\baselineskip=12pt \rm}%
\def\smalltype{%
\let\rm=\eightrm \let\bf=\eightbf%
\let\it=\eightit \let\sl=\eightsl%
\let\tt=\eighttt
\def\cal{\fam2 }%
 \textfont0=\eightrm   \scriptfont0=\sevenrm \scriptscriptfont0=\sixrm
 \textfont1=\eighti    \scriptfont1=\seveni  \scriptscriptfont1=\sixi
 \textfont2=\eightsy   \scriptfont2=\sevensy \scriptscriptfont2=\sixsy
 \textfont3=\tenex     \scriptfont3=\tenex   \scriptscriptfont3=\tenex
 \textfont\itfam=\eightit
 \textfont\slfam=\eightsl
 \textfont\bffam=\eightbf
 \textfont\ttfam=\eighttt
\baselineskip=10pt
\parskip=1pt plus1pt minus 1pt%
\rm}%
%--------------------------------------------------------------------------
%       Spacing
%--------------------------------------------------------------------------
\voffset=0truein%
\vsize=9.3truein%
%\hoffset=0.35truein%
\hoffset=0.1truein%
\hsize=6.0truein%
\parindent=0.375truein%
\widowpenalty=300

\def\spaceandahalf{%
\baselineskip=15pt% spacing and a half
\parskip=4pt plus4pt minus2pt%
\everydisplay={\openup 1.5\jot}}%

\def\singlespace{%
\baselineskip=12pt% single spacing
\parskip=1pt plus1pt minus 1pt%
\everydisplay={\openup 0\jot}}%

\def\closeup{\singlespace}%

%TeXbook page 416
\newdimen\pagewidth
\newdimen\pageheight    
\pagewidth=6.0truein
\vsize=9.3truein%
\pageheight=9.3truein
%--------------------------------------------------------------------------
%       newifs and stuff
%--------------------------------------------------------------------------
\newif\ifwindowcover%
\windowcoverfalse%
\outer\def\windowcover{\windowcovertrue}%
\newif\ifcam%
\camfalse%
\outer\def\damtpaddress{\camtrue}%
\newif\ifnofigs%
\nofigsfalse%
\newif\ifchapstart
\chapstartfalse
\def\chapstartpage{\global\chapstarttrue}
\newif\ifinappendices%
\inappendicesfalse%
\newif\iffiguresatend%
\newif\ifnumbereqnsseq%
\def\numbereqnsseq{ \numbereqnsseqtrue}%
\newif\ifnumberfigsseq%
\def\numberfigsseq{ \numberfigsseqtrue}%
\newif\ifnumbertablesseq%
\def\numbertablesseq{ \numbertablesseqtrue}%
\newif\ifnumberrefsintext%
\def\numberrefsintext{ \numberrefsintexttrue}%
\newif\ifinrefs%
\inrefsfalse%
\newif\ifpaperstyle%
\outer\def\paperstyle{\paperstyletrue
     \refauthorinitialssurname%
     \numbereqnsseq \numberfigsseq \numbertablesseq%
     \numberrefsintext}%
\newif\ifsendstyle%
\newif\ifnyasrefstyle%

\newif\ifsinglesided%
\newif\ifdated%

\newif\ifsastyle%
\outer\def\sastyle{\sastyletrue\numberrefsintext%
\ifsinglesided\headline={\rightheadline}
\else\headline={\ifodd\pageno\rightheadline
                        \else\leftheadline\fi}\fi%
\footline={\ifdated\hfil\eightrm\today\else{}\fi}}
\newif\ifheadset%
\headsetfalse%

\countdef\sectno=1%
\countdef\subsectno=2%
\countdef\subsubsectno=3%
\newcount \equano%
\newcount \figno%
\newcount \tableno%
\newcount \referencecount%
\newbox\TitlePage%
\newbox\DedicationPage%
\newbox\PrefacePage%
\newbox\ConclusionPage%
\newbox\AcknowledgementsPage%
\newbox\SummaryPage%
\newbox\TableOfContents%
\newbox\ListOfFigures%
\newbox\Nomenclature%
\newbox\ListOfTables%
\newbox\FiguresPage%
\newbox\figurebox%
\def\thesectno{%
\ifinappendices%
  \ifcase\sectno\or%
    A\or B\or C\or D\or E\or F\or G\or H\or I\or J\or K\or L\or M%
    N\or O\or P\or Q\or R\or S\or T\or U\or V\or W\or X\or Y\or Z\fi%
\else%
  \the\sectno%
\fi}%
\def \lastsubsectno{\thesectno .\the\subsectno}%
\def \lastsubsubsectno{\thesectno .\the\subsectno .\the\subsubsectno}%
\def \lasteqno{\ifnumbereqnsseq (\the\equano)%
\else (\thesectno.\the\equano)\fi}%
\def \preveqno{\advance \equano by -1 \lasteqno}%
\def \lastfigno{\ifnumberfigsseq \hbox{Figure}~\hbox{\the\figno}%
\else \hbox{Figure}~\hbox{\thesectno -\the\figno}\fi}
\def \lasttableno{\ifnumbertablesseq \hbox{Table}~\hbox{\the\tableno}%
\else \hbox{Table}~\hbox{\thesectno .\the\tableno}\fi}%
\def \nextsubsectno{{\advance \subsectno by 1 \lastsubsectno}}%do this locally
\def \nextsubsubsectno{{\advance \subsubsectno by 1 \lastsubsubsectno}}%
\def \nexteqno{{\advance \equano by 1 \lasteqno}}%
\def \nextfigno{{\advance \figno by 1 \lastfigno}}%
\def \nextnextfigno{{\advance \figno by 2 \lastfigno}}%
\def \nexttableno{{\advance \tableno by 1 \lasttableno}}%
\def\minimalsectno{\ifnum\subsubsectno>0 \lastsubsubsectno%
\else\ifnum\subsectno>0 \lastsubsectno%
\else \thesectno \fi\fi}%
%--------------------------------------------------------------------------
%       Headings for start of sections
%--------------------------------------------------------------------------
\def\beginsect#1\par{\begingroup
\ifsastyle\thesisheading{#1}
        \chapstartpage
%    \accretetoTOCone{\ifinrefs\else\thesectno.\fi}{#1}%
    \xdef\rhead{\ #1}
\else%
    \global \advance \sectno by 1%
    \mainheading{#1}%
%    \accretetoTOCone{\ifinrefs\else\thesectno.\fi}{#1}%
\fi
\global \subsectno=0 \global \subsubsectno=0%
\ifnumbereqnsseq \else \global \equano=0 \fi%
\ifnumberfigsseq \else \global \figno=0 \fi%
\ifnumbertablesseq \else \global \tableno=0 \fi%
\message{\thesectno . #1}%
\nobreak\endgroup%
\ifsastyle\headsetfalse\else\fi}%

\def\saheading#1{\vfill\supereject
        \global \advance \sectno by 1%
        \chapstartpage
        \vskip 0pt plus.3\vsize \penalty-100 \vskip 0pt plus-.3\vsize%
        \ifinrefs\noindent{\bigtype \bf #1}
%       \else\ifinappendices\noindent{\bigtype \bf Appendix \thesectno
%.\ #1}
        \else\ifinappendices\noindent{\bigtype \bf #1}
        \else\noindent{\bigtype \the\sectno .\ \bf #1}\fi\fi
        \par\nobreak\noindent}

\def\thesisheading#1{\vfill\supereject
        \global \advance \sectno by 1%
        \vskip 0pt plus.3\vsize \penalty-100 \vskip 0pt plus-.3\vsize%
        \ifinrefs\noindent{\bigtype \bf #1}
        \else\ifinappendices\noindent{\bigtype \bf Appendix \thesectno .\ #1}
        \else\line{}\vskip 0.5truein
        \centerline{\twelverm CHAPTER \the\sectno}
                \vskip 0.5truein
        \centerline{\fourteenbf #1}
        \vskip 0.5truein
        \fi\fi
        \par\nobreak\noindent}

\def\mainheading#1{\begingroup\nohyphens\hbadness=3000%
        \ifinrefs%
          \noindent {\bf #1}\nobreak%
        \else%
           \ifinappendices%
              \noindent {\bf #1}\par\nobreak%
           \else
              \noindent {\bf \thesectno . #1}\par\nobreak%
           \fi%
        \fi%
    \noindent\endgroup}%

\def\subheading#1{\begingroup\nohyphens\hbadness=3000%
%  \vskip 0pt plus.2\vsize \penalty-100 \vskip 0pt plus-.2\vsize%
%\medskip\vskip\parskip%
\bigskip
    \noindent {\sl\bf\lastsubsectno . #1}\par\nobreak\noindent%
\endgroup}%

\def\subsubheading#1{\begingroup\nohyphens\hbadness=3000%
  \vskip 0pt plus.1\vsize \penalty-100 \vskip 0pt plus-.1\vsize%
\medskip\vskip\parskip%
  \noindent {\sl\lastsubsubsectno . #1}\par\nobreak\smallskip\noindent%
\endgroup}%

\outer \def \beginappendix#1\par{%
\inrefsfalse%
\ifinappendices\else\inappendicestrue\global\sectno=0\fi%
\begingroup%
\ifsastyle\saheading{#1}%
\xdef\rhead{#1}
\else%
    \global \advance \sectno by 1%
    \mainheading{#1}%
    \ifnumbereqnsseq \else \global \equano=0 \fi%
    \ifnumberfigsseq \else \global \figno=0 \fi%
    \ifnumbertablesseq \else \global \tableno=0 \fi%
    \global \subsectno=0 \global \subsubsectno=0%
\fi%
\message{\thesectno . #1}%
%\accretetoTOCone{}{Appendix \thesectno - #1}%
%\accretetoTOCone{}{#1}%
\endgroup%
\ifsastyle\headsetfalse\else\fi%
}

\outer \def \beginsubsect#1\par{
\global \advance \subsectno by 1%
\global \subsubsectno=0%
\accretetoTOCtwo{\lastsubsectno.}{#1}%
\subheading{#1}}

\outer \def \beginsubsubsect#1\par{\global \advance \subsubsectno by 1%
\accretetoTOCthree{\lastsubsubsectno.}{#1}\subsubheading{#1}}%
%--------------------------------------------------------------------------
%       labels for equations and other things
%--------------------------------------------------------------------------
\def \myeqno{\global \advance \equano by 1 {\rm \lasteqno}}%
\def \myeqlabel#1{\expandafter\xdef\csname #1\endcsname{\lasteqno}}

\def \mytextlabel#1{\expandafter\xdef\csname #1\endcsname{\minimalsectno}}
\def \myChapterlabel#1#2{%
 \expandafter\xdef\csname #1\endcsname{Chapter~#2}}
\def \myAppendixlabel#1#2{%
 \expandafter\xdef\csname #1\endcsname{Appendix~\hbox{#2}}}
\def \myfigurelabel#1{\expandafter\xdef\csname #1\endcsname{\lastfigno}}
\def \mytablelabel#1{\expandafter\xdef\csname #1\endcsname{\lasttableno}}
%--------------------------------------------------------------------------
%       Figures
%--------------------------------------------------------------------------
\input epsf

\def\figure #1 #2 #3\par{\global \advance \figno by 1%
\iffiguresatend%
%accrete to the List Of Figures and FiguresPage
   \global\setbox\ListOfFigures=\vbox{\unvbox\ListOfFigures%
   \bigskip\caption{\hbox{\bf\lastfigno. }{#3}}
   \par\filbreak}%
   \ifnofigs%
   \else
      \global\setbox\FiguresPage=\vbox{\unvbox\FiguresPage%
        \epsfysize=#1
        \centerline{\epsffile{#2}}
       \vskip 0.5truein\relax%
       \hbox{\bf \lastfigno}
        {#3}
        \vfill\eject}
   \fi%
\else%
\ifsastyle
   \global\setbox\ListOfFigures=\vbox{\unvbox\ListOfFigures%
   \bigskip\caption{\hbox%
{\bf\lastfigno. }
{page \folio},
{size in text #1},
{file #2}\filbreak\noindent{#3}}
   \par\filbreak}%
\fi
  \setbox\figurebox=\vbox{%
  \ifnofigs
  \else
        \epsfysize=#1   
        \centerline{\epsffile{#2}}
  \fi
    \caption{{\bf\lastfigno. }{\eightsl #3}}}
  \topinsert\unvbox\figurebox\endinsert%
\fi}%

\def\twofigures #1 #2 #3 #4\par{\global \advance \figno by 1%
\iffiguresatend %accrete to the List Of Figures and FiguresPage
   \begingroup%
   \global\setbox\ListOfFigures=\vbox{\unvbox\ListOfFigures%
   \bigskip\caption{\hbox{\bf\lastfigno. }{#4}}
   \par\filbreak}%
   \ifnofigs
   \else
      \global\setbox\FiguresPage=\vbox{\unvbox\FiguresPage%
        \epsfysize=#1
        \centerline{\epsffile{#2}\hfil\epsffile{#3}}
       \vskip 0.5truein\relax%
       \hbox{\bf \lastfigno}
        \vfill\eject}
   \fi
   \endgroup%
\else%
  \setbox\figurebox=\vbox{%
  \ifnofigs
  \else
        \centerline{
        \epsfysize=#1   
        (a)\epsffile{#2}
        \ \ 
        \epsfysize=#1   
        (b)\epsffile{#3}}
  \fi
    \caption{{\bf\lastfigno. }{\eightsl #4}}}
  \topinsert\unvbox\figurebox\endinsert%
\fi}%

\def\fourfigures #1 #2 #3 #4 #5 #6\par{\global \advance \figno by 1%
\iffiguresatend %accrete to the List Of Figures and FiguresPage
   \begingroup%
   \global\setbox\ListOfFigures=\vbox{\unvbox\ListOfFigures%
   \bigskip\caption{\hbox{\bf\lastfigno. }{#6}}
   \par\filbreak}%
   \ifnofigs
   \else
      \global\setbox\FiguresPage=\vbox{\unvbox\FiguresPage%
        \centerline{
        \epsfysize=#1   
        (a)\epsffile{#2}
        \ \ 
        \epsfysize=#1   
        (b)\epsffile{#3}}
        \line{ }
        \centerline{
        \epsfysize=#1   
        (c)\epsffile{#4}
        \ \ 
        \epsfysize=#1   
        (d)\epsffile{#5}}
        \hbox{\bf \lastfigno}
         {#6}
        \vfill\eject}
   \fi
   \endgroup%
\else%
  \setbox\figurebox=\vbox{%
  \ifnofigs
  \else
        \centerline{
        \epsfysize=#1   
        (a)\epsffile{#2}
        \ \ 
        \epsfysize=#1   
        (b)\epsffile{#3}}
        \line{ }
        \centerline{
        \epsfysize=#1   
        (c)\epsffile{#4}
        \ \ 
        \epsfysize=#1   
        (d)\epsffile{#5}}
  \fi
    \caption{{\bf\lastfigno. }{\eightsl #6}}}
  \topinsert\unvbox\figurebox\endinsert%
\fi}%

\def\eightfigures #1 #2 #3 #4 #5 #6 #7 #8 #9%
\par{\global \advance \figno by 1%
\iffiguresatend %accrete to the List Of Figures and FiguresPage
   \begingroup%
   \global\setbox\ListOfFigures=\vbox{\unvbox\ListOfFigures%
   \bigskip\caption{\hbox{\bf\lastfigno. }{#6}}
   \par\filbreak}%
   \ifnofigs
   \else
      \global\setbox\FiguresPage=\vbox{\unvbox\FiguresPage%
        \epsfysize=#1
        \centerline{\epsffile{#2}\hfil\epsffile{#3}}
       \vskip 0.5truein\relax%
       \hbox{\bf \lastfigno}
        \vfill\eject}
   \fi
   \endgroup%
\else%
  \setbox\figurebox=\vbox{%
  \ifnofigs
  \else
        \centerline{
        \epsfysize=1.8truein
        (a)\epsffile{#1}
        \epsfysize=1.8truein
        (b)\epsffile{#2}
        \epsfysize=1.8truein
        (c)\epsffile{#3}}
        \line{ }
        \centerline{
        \epsfysize=1.8truein
        (d)\epsffile{#4}
        \epsfysize=1.8truein
        (e)\epsffile{#5}
        \epsfysize=1.8truein
        (f)\epsffile{#6}}
        \line{ }
        \centerline{
        \epsfysize=1.8truein
        (g)\epsffile{#7}
        \epsfysize=1.8truein
        (h)\epsffile{#8}
        \epsfysize=1.8truein
        \ \ \ \epsffile{key.eps}}
  \fi
    \caption{{\bf\lastfigno. }{\eightsl #9}}
}
  \topinsert\unvbox\figurebox\endinsert%
\fi}%

%see p.355 and p.356
\def\caption#1{%
\smallskip%
\begingroup% keep the changes local.
\nohyphens%
\advance \leftskip by \parindent%
\advance \rightskip by \parindent% plus4em%
\spaceskip=.3333em plus .3333em \xspaceskip=.5em plus .5em%
\par \noindent{%
\iffiguresatend{#1}\else\smalltype{#1}\par\fi\par} \par\endgroup}%

%--------------------------------------------------------------------------
%       References
%--------------------------------------------------------------------------
\outer \def \beginrefs{\inrefstrue\inappendicesfalse}%

\def\showrefs{\ifsastyle\accretetoTOCone{}{References}\headsetfalse\else\fi%
\setbox\myrefs=\vbox{\unvbox\myrefs\vfil}%
\ifsendstyle\vfill\eject\else\smallskip\goodbreak\fi
    \unvbox\myrefs}

\def\addtorefs#1#2{%
\global\advance\referencecount by 1%
\expandafter\xdef\csname#2\endcsname{\the\referencecount}%
\global\setbox\myrefs=\vbox{\unvbox\myrefs%
{\ifsendstyle\else\closeup\fi
\ifnyasrefstyle
\item{\csname#2\endcsname.}\frenchspacing{\csname#1\endcsname}
\else
\item{[\csname#2\endcsname]}\frenchspacing{\csname#1\endcsname}
\fi
\hfil\par\ifsendstyle\medskip\fi}}}

\def\addref#1#2{\ifundefined{#2}\addtorefs{#1}{#2}\fi
\ifnyasrefstyle$^{\csname#2\endcsname}$\else~[\csname#2\endcsname]\fi}%

% don't show this in text

\def\addtworefs#1#2#3#4{%
{\ifundefined{#2}\addtorefs{#1}{#2}\fi}%
{\ifundefined{#4}\addtorefs{#3}{#4}\fi}%
\ifnyasrefstyle
$^{\csname#2\endcsname,~\csname#4\endcsname}$%
\else~[\csname#2\endcsname,~\csname#4\endcsname]\fi}%

\def\addthreerefs#1#2#3#4#5#6{% 
{\ifundefined{#2}\addtorefs{#1}{#2}\fi}%
{\ifundefined{#4}\addtorefs{#3}{#4}\fi}%
{\ifundefined{#6}\addtorefs{#5}{#6}\fi}%
\ifnyasrefstyle
%$^{\csname#2\endcsname,~\csname#4\endcsname,~\csname#6\endcsname}$%
%\else~[\csname#2\endcsname,~\csname#4\endcsname,~\csname#6\endcsname]\fi}%
$^{\csname#2\endcsname,\csname#4\endcsname,\csname#6\endcsname}$%
\else~[\csname#2\endcsname,\csname#4\endcsname,\csname#6\endcsname]\fi}%

\def\addfourrefs#1#2#3#4#5#6#7#8{% none should have been refd before
{\ifundefined{#2}\addtorefs{#1}{#2}\fi}%
{\ifundefined{#4}\addtorefs{#3}{#4}\fi}%
{\ifundefined{#6}\addtorefs{#5}{#6}\fi}%
{\ifundefined{#8}\addtorefs{#7}{#8}\fi}%
\ifnyasrefstyle
%$^{\csname#2\endcsname,~\csname#4\endcsname,~\csname#6\endcsname}$%
%\else~[\csname#2\endcsname,~\csname#4\endcsname,~\csname#6\endcsname]\fi}%
$^{\csname#2\endcsname-\csname#8\endcsname}$%
\else~[\csname#2\endcsname-\csname#8\endcsname]\fi}%

% in the text of the article:
%   use \addref{refA1}{A1} to include reference with index A1
% in the reference file:
%\refarticle {refA1}{A.U. Thor1, A.U. Thor2 and A.U. Thor3}{year}
%                   {Title with only the first word capitalized}
%                   {J.}{volume}{firstpage--lastpage}
%\refbook    {refA1}{A.U. Thor1, A.U. Thor2 and A.U. Thor3}{year}
%                   {Title with all words capitalized}
%                   {Publisher}{Location}
%\refartbook {refA1}{A.U. Thor1, A.U. Thor2 and A.U. Thor3}{year}
%                   {Title of the article with only the first word capitalized}
%                   {Title of the book    with all words capitalized}
%                   {E.D. Tor1 and E.D. Tor2}  (don't put the (eds) in)
%                   {Publisher}{Location}{000--999}
%\refphd     {refA1}{A.U. Thor}{year}
%                   {Title with only the first word capitalized}
%                   {University}

\def\genericref#1#2{\begingroup\closeup%
\expandafter\xdef\csname#1\endcsname{#2.}%
\endgroup}%
\def\refyear#1{\ifnyasrefstyle #1. \else (#1)\fi}%
\def\reftitle#1{\lq\lq#1''}%
\def\refauthorinitialssurname{%
\def\author##1##2{##1~##2}
\def\authorn##1##2{\ifnyasrefstyle\uppercase{##2,~##1}\else{##1~##2}\fi}}%
\refauthorinitialssurname%
\outer\def\refarticle#1#2#3#4#5#6#7{%
\genericref{#1}{%
\ifnyasrefstyle
    {#2.\ \refyear{#3} \reftitle{#4}. {#5} {\bf #6}: {#7}}%
\else%
  \ifpaperstyle%
%    {#2,\ {#5}~{\bf{#6}},\ \hbox{#7} \ \refyear{#3}}%
    {#2\ \reftitle{#4.} {\sl #5} {\bf #6} #7\ \refyear{#3}}%
  \else
    {#2\ \reftitle{#4.} {\sl #5} {\bf #6} #7\ \refyear{#3}}%
  \fi%  
\fi}}%

\outer\def\refarticletobe#1#2#3#4#5{%
\genericref{#1}{#2, \reftitle{#4} {#5} \refyear{#3}}}

\outer\def\refbook#1#2#3#4#5#6{%
\genericref{#1}%
{\ifnyasrefstyle%
#2.\ #3. #4. #5, #6%
\else%
\ifpaperstyle%
{#2,\ {\sl #4}\ (#5, \ #6, \ #3)}%
  \else%
    {#2\ {\sl #4}.  #5: #6 \ \refyear{#3}}%
\fi\fi}}%

\outer\def\refartbook#1#2#3#4#5#6#7#8#9{%
\genericref{#1}%
{\ifpaperstyle%
%  {#2, in {\sl #5}, edited by #6 (#7, #8, #3), pp#9}%
    {#2 \refyear{#3} \reftitle{#4,} in {\sl #5}, #6 (Ed.). #7: #8}%
\else%
    {#2 \reftitle{#4} #9 in {\sl #5}, Ed. #6. #7: #8 \refyear{#3}}%
\fi}}%

\outer\def\refphd#1#2#3#4#5{%
\genericref{#1}%
{\ifpaperstyle%
%  {#2, #4 (#5, Ph.D. thesis, #3)}%
  {#2, {\sl #4} (Ph.D. thesis, #5, #3)}%
\else%
    {#2 \refyear{#3} \reftitle{#4.} {#5, Ph.D. thesis}}%
\fi}}

\outer\def\refnote#1{%
\genericref{#1}}
%--------------------------------------------------------------------------
%       Title page etc
%--------------------------------------------------------------------------
\def\theauthor{Grant Lythe}

\def\title#1\par{\message{#1}%
\xdef\lhead{#1}%
\xdef\rhead{}%
\begingroup%accrete to the Title Page
  \ifwindowcover%
    \dimen1=2.5truein \advance\dimen1 by -\voffset%
    \global\setbox\TitlePage=\vbox{\vbox to \dimen1{%
      \dimen2=1.75truein \advance\dimen2 by -\voffset%
      \smallskip \relax%
      \dimen3=1.5truein \advance\dimen3 by -\hoffset%
      \leftskip=\dimen3 plus3em\relax%
      \dimen4=\hsize \advance\dimen4 by \hoffset \advance\dimen4 by -5truein%
      \rightskip=\dimen4 plus3em\relax%
      \parfillskip=0pt \parindent=0pt \spaceskip=0.3333em
        \xspaceskip=0.5em%
        \line{}\bigskip\line{}\bigskip
      \centerline{\bigtype\bf #1}\smallskip
          {\centerline{\theauthor}}
        }}
  \fi%
\endgroup}%

\def\signaturepaper{
\global\setbox\TitlePage=\vbox{%
\unvbox\TitlePage%
\noindent\narrower\raggedcenter{\bigtype%
\ifwindowcover \else \theauthor\par\fi%
\ifcam\damtp\else\ont\fi
\bigskip}}}%

\def\signaturethesis{%
\parskip=20pt plus 10pt%
\global\setbox\TitlePage=\vbox{%
\line{}\bigskip
\vfill%
\noindent\narrower\raggedcenter{%
\bigtype%
{\bf Stochastic slow-fast dynamics}\par%
\bigtype%
Grant David Lythe\par%
\medtype
Trinity College, Cambridge\par%
\vfill%
Dissertation submitted\break%
for the degree of\break%
Doctor of Philosophy\break%
at the\break%
University of Cambridge\break%
\vfill%
October 1994}}}%

\def\signature{\begingroup%
\ifpaperstyle \signaturepaper \fi%
%\ifsastyle \signaturesa \fi%
\ifsastyle \signaturethesis \fi%
\endgroup}%

\long\def\addtotitlepage#1{%accrete to the Title Page
\global\setbox\TitlePage=\vbox{%
\unvbox\TitlePage%
\bigskip\noindent#1\par}}%

\def\abstract#1\par{\begingroup%
\addtotitlepage{{\centerline{\bf Abstract}}\par\nobreak\medskip\noindent#1}%
\endgroup}%

\def\submittedto#1{\begingroup\closeup%
\addtotitlepage{\smalltype
\ifpaperstyle  {Submitted to  \hfill printed on \ \today}\fi%
}\endgroup}%

\def\publishedin#1{\begingroup\closeup%
\addtotitlepage{\bigtype
\ifpaperstyle  {#1}\fi}\endgroup}%

\long\def\summary#1{%
\global\setbox\SummaryPage=\vbox{%
  \unvbox\SummaryPage \noindent #1\par}%
  \begingroup\pageno=-4\endgroup}

\outer \def \beginack{
\noindent{\bf{Acknowledgement}}\break}

\def\Showbox#1{\begingroup%
\loop%
\ifvbox#1%
    \dimen255=\vsize%
    \advance \dimen255 by -\topskip%
    \setbox0=\vsplit#1 to \dimen255%
    \unvbox0%
    \null \vfill \penalty -10000%
\repeat%
\endgroup}%

\def\ShowTP{\begingroup \nopagenumbers\headline={}%
\ifvbox\TitlePage%
    \global \sectno=0 \global \subsectno=0 \global \subsubsectno=0%
    \setbox\TitlePage=\vbox{\unvbox\TitlePage\vfil}%
    \Showbox\TitlePage%
\fi \endgroup}%

\def\PreambleHeader#1{%
\begingroup\raggedcenter\line{}\vskip 0.3truein 
\ifsastyle\bigtype\else\Bigtype\fi\bf \noindent%
#1\par \vskip 0.2truein\endgroup\noindent}%

\def\ShowAck{%
\ifvbox\AcknowledgementsPage%
    \global \sectno=0 \global \subsectno=0 \global \subsubsectno=0%
    \setbox\AcknowledgementsPage=\vbox{%
\PreambleHeader{Acknowledgements}\unvbox\AcknowledgementsPage \vfil}%
    \Showbox\AcknowledgementsPage%
\fi}%

\def\ShowSummary{%
\ifvbox\SummaryPage%
    \global \sectno=0 \global \subsectno=0 \global \subsubsectno=0%
    \setbox\SummaryPage=\vbox{%
\PreambleHeader{Summary}\unvbox\SummaryPage \vfil}%
    \Showbox\SummaryPage%
\fi}%

\def\Trailers{%
\pageno=-1%
\footline={}
\ifsastyle\headline={}%
%\ShowDed\ShowPref\ShowSummary
 \ShowTOC \fi
\iffiguresatend \begingroup \nopagenumbers \ShowLOF \ShowFP \endgroup\fi%
}%
%--------------------------------------------------------------------------
% Useful bits and pieces 
%--------------------------------------------------------------------------
% see p.406
\def\themonth{\ifcase\month\or%
  January\or February\or March\or April\or%
  May\or June\or July\or August\or%
  September\or October\or November\or December\fi}%
\def\today{\number\day%
\space%
\themonth%
\space%
\number\year}%
%
%see p. 356 and p. 98
\outer\def\raggedright{\rightskip=0pt plus 4em%
  \spaceskip=.3333em \xspaceskip=.5em%
  \pretolerance=200 \tolerance=400}%
% from p. 223
\def\leaderfill{\leaders\hbox to 1em{\hss.\hss}\hfill}%
% from p. 359 and p. 81
\def\dddot#1{\begingroup\vbox{%
\moveleft 0.15em\hbox{...}%
\nointerlineskip\vskip0.4ex%
\hbox{$\displaystyle{#1}$}}\endgroup}%
%--------------------------------------------------------------------------
% Bits and pieces written by Alastair
%--------------------------------------------------------------------------
%from p. 308
\def\ifundefined#1{\expandafter\ifx\csname#1\endcsname\relax}%
%special hyphenations see p. 452
\hyphenation{co-dimen-sion sol-ution sol-utions sun-spot sun-spots
pro-pen-sity}
%from p. 348 and 307 -- get rid over the overfull rule.
\overfullrule=0pt%
%from excercise 14.34
\def\nohyphens{\pretolerance=9999 \tolerance=9999%
\hyphenpenalty=9999 \exhyphenpenalty=1000 }%
\def\raggedcenter{\leftskip=0pt plus6em \rightskip=\leftskip%
\parfillskip=0pt \parindent=0pt \spaceskip=0.3333em \xspaceskip=0.5em%
\nohyphens \hbadness=3000}%
% Tables:
% example: \topinsert
%          table material goes here
%          \table table caption goes here
%
%          \endinsert
% sample table material:
%   \line{\hfil\vbox{\openup 1\jot\hrule\smallskip
%                    \halign{<table stuff -- see p. 235>}
%                    \smallskip\hrule}\hfil}
%
% a useful preamble for several aligned equations (more than one per line)
%  see p.242 of the TeXbook
% $$\vcenter{\openup 1\jot \halign{
%        $\displaystyle\hfil#$&$\displaystyle{}#\hfil$
% &&\quad$\displaystyle\hfil#$&$\displaystyle{}#\hfil$\cr
% R&=R1 & Q&=Q1 & \cr etc
%
% a useful preamble for the cases type table
%  see p.242 of the TeXbook
% \topinsert
% \line{\hfil\vbox{\openup 1\jot\hrule\smallskip \halign{
% \span\generaltablepreamble
% T & $expression$ & $expression$ \cr etc
%
% or better:
% \quad\hfil#\hfil\quad&&\hfil#\hfil\quad\cr
%
\outer\def \table#1\par{\global \advance \tableno by 1%
\begingroup \closeup%accrete to the List Of Tables%
\global\setbox\ListOfTables=\vbox{\unvbox\ListOfTables%
\smallskip\noindent{\bf\lasttableno.\ }{\strut\sl #1\strut}\par}\endgroup%
\medskip\caption{{\bf\lasttableno.\ }{\sl #1}}}%

\paperstyle

\newbox\myrefs%
    \setbox\myrefs=\vbox{%
\ifsastyle\noindent{\bigtype \bf{References}\medskip}
\else\smallskip\noindent\bf References\par\medskip\fi}%

\beginrefs 
\begingroup\raggedright%
\nohyphens%
%\input mssymb

%A11
\refarticle{refA11}
{\author{V.S.}{Afra\u\i movich}, \author{V.V.}{Bykov} \and
 \author{L.P.}{Shil'nikov}}{1983}
{On structurally unstable attracting limit sets of Lorenz attractor type}
            {Trans. Moscow Math. Soc.}{44}{153--216}
 
%A19
\refartbook{refA19}
{\author{V.S.}{Afra\u\i movich} \and \author{L.P.}{Shil'nikov}}{1983}
            {Strange attractors and quasiattractors}
            {Nonlinear Dynamics and Turbulence}
            {\author{G.I.}{Barenblatt}, \author{G.}{Iooss} \and
             \author{D.D.}{Joseph}}
            {Pitman}{Boston}{1--34}
 
%A18
\refarticle{refA18}
{\author{G.}{Ahlers} \and \author{I.}{Rehberg}}{1986}
            {Convection in a binary mixture heated from below}
            {Phys. Rev. Lett.}{56}{1373--1376}
 
%A21
\refarticle{refA21}
{\author{K.H.}{Alfsen} \and \author{J.}{Fr\o yland}}{1985}
            {Systematics of the Lorenz model at $\sigma=10$}
            {Physica Scripta}{31}{15--20}
 
%A22
\refarticle{refA22}
{\author{I.S.}{Aranson}, \author{V.S.}{Afra\u\i movich} \and
 \author{M.I.}{Rabinovich}}{1989}
            {Multidimensional strange attractors and turbulence}
            {Sov. Sci. Rev.~C. Math. Phys.}{8}{293--376}
 
%A10
\refarticle{refA10}
{\author{A.}{Arn\'eodo}, \author{P.}{Coullet} \and \author{C.}{Tresser}}{1981}
            {A possible new mechanism for the onset of turbulence}
            {Phys. Lett.}{81A}{197--201}
 
%A4
\refarticle{refA4}
{\author{A.}{Arn\'eodo}, \author{P.H.}{Coullet} \and \author{E.A.}{Spiegel}}{1982}
            {Chaos in a finite macroscopic system}
            {Phys. Lett.}{92A}{369--373}
 
%A2
\refarticle{refA2}
{\author{A.}{Arn\'eodo}, \author{P.H.}{Coullet} \and \author{E.A.}{Spiegel}}
            {1985}
            {The dynamics of triple convection}
            {Geophys. Astrophys. Fluid Dynamics}{31}{1--48}
 
%A3
\refarticle{refA3}
{\author{A.}{Arn\'eodo}, \author{P.H.}{Coullet}, \author{E.A.}{Spiegel} \and
 \author{C.}{Tresser}}
            {1985}
            {Asymptotic chaos}
            {Physica}{14D}{327--347}
 
%A13
\refarticle{refA13}
{\author{A.}{Arn\'eodo} \and \author{O.}{Thual}}{1985}
            {Direct numerical simulations of a triple convection problem
             versus normal form predictions}
            {Phys. Lett.}{109A}{367--373}
 
%A15
\refarticle{refA15}
{\author{V.I.}{Arnol'd}}{1972}
           {Lectures on bifurcations in versal families}
           {Russ. Math. Surveys}{27}{54--123}
 
%A14
\refbook{refA14}
{\author{V.I.}{Arnol'd}}{1983}
            {Geometrical Methods in the Theory of Ordinary Differential
             Equations}
            {Springer}{New York}
 
%A9
\refarticle{refA9}
{\author{W.}{Arter}}{1983}
            {Nonlinear convection in an imposed horizontal magnetic field}
            {Geophys. Astrophys. Fluid Dynamics}{25}{259--292}
 
%A100
\refbook{refA100}
{\author{L.}{Arnold}}{1974}
    {Stochastic Differential Equations}
    {Wiley}{New York}

%A101
\refbook{refA101}
{\author{Milton}{Abramowitz} \and \author{Irene}{Stegun}}{1968}
    {Handbook of Mathematical Functions}
    {Dover}{New York}
 
%A102
\refarticle{refA102}
{\author{*}{McA**}}{1888}
           {Lognormal stuff}
           {J.~Appl. Math. Phys. (ZAMP)}{37}{608--623}

%A103
\refarticle{refA103}
{\author{Guenter}{Ahlers}, \author{Christopher W.}{Meyer}
\& \author{I.}{Rehberg}}{1989}
            {Deterministic and Stochastic Effects Near the Convective Onset}
            {J. Stat. Phys.}{54}{1121--1131}
%A104
\refarticle{refA104}
{\author{Guenter}{Ahlers}, \author{M.C.}{Cross}, \author{P.C.}{Hohenberg}
\and \author{S.}{Safran}}{1981}
            {The amplitude equation near the convective threshold:
application to time-dependent heating experiments}
            {J. Fluid Mech.}{110}{297--334}

%A105
\refbook{refA105}
{\author{V.I.}{Arnol'd (Ed.)}}{1994}
            {Dynamical Systems V}
            {Springer}{Berlin}
%A106
\refbook{refA106}
{\author{Robert J.}{Adler}}{1981}
            {The Geometry of Random Fields}
            {Wiley}{Chichester}
%A107
\refbook{refA107}
{\author{Robert J.}{Adler}}{1990}
            {An Introduction to Continuity, Extrema and Related
Topics for General Gaussian Processes}
            {Institute of Mathematical Statistics}{Hayward, California}

\refarticle{refA108}
{\author{Dieter}{Armbruster}, \author{John}{Guckenheimer}
 \and \author{Philip}{Holmes}}{1988}
           {Heteroclinic cycles and modulated travelling waves in
systems with O(2) symmetry}
           {Physica D}{29}{257--282}
 
%B2
\refbook{refB2}
{\author{P.}{Berg\'e}, \author{Y.}{Pomeau} \and \author{C.}{Vidal}}{1984}
            {Order Within Chaos}
            {Wiley}{New York}

\refarticle{refB3}
{\author{C.S.}{Bretherton} \and \author{E.A.}{Spiegel}}{1983}
           {Intermittency through modulational instability}
           {Phys. Lett.}{96A}{152--156}
 
%B25
\refarticle{refB25}
{\author{F.H.}{Busse} \and \author{A.C.}{Or}}{1986}
           {Subharmonic and asymmetric convection rolls}
           {J.~Appl. Math. Phys. (ZAMP)}{37}{608--623}
 
%B100
\refarticle{refB100}
{\author{G.}{Broggi}, \author{A.}{Colombo}, \author{L.A.}{Lugiato}
\and \author{Paul}{Mandel}}{1986}
          {Influence of white noise on delayed bifurcations}
           {Phys. Rev. A}{33}{3635--3637}

\refarticle{refB101}
{\author{C.}{van den Broeck} \and \author{Paul}{Mandel}}{1987}
           {Delayed bifurcations in the presence of noise}
           {Phys. Lett.}{A122}{36--38}

\refarticle{refB102}
{\author{Jean-Marie}{Wersinger}, \author{John M.}{Finn}
  \and \author{Edward}{Ott}}{1980}
            {Bifurcations and strange behaviour in instability
saturation by nonlinear mode coupling}
         {Phys. Rev. Lett.}{44}{453--457}

\refarticle{refB103}
{\author{J.Robert}{Buchler}, \author{Pawe\l}{Moskalik} \and\author{G\'eza}{Kov\'acs}}{1991}
           {Periodic stellar pulsations: stability analysis and
amplitude equations}
           {Ap. J.}{380}{185--199}

\refarticle{refB104}
{\author{J.R.}{Buchler} \and\author{G.}{Kov\'acs}}{1986}
            {The effects of a 2:1 resonance in nonlinear radial
stellar pulsations}
            {Ap. J.}{303}{749--765}

%B105
\refartbook{refB105}
{\author{Claude}{Baesens}}{1991}
{Noise effect on dynamic bifurcations: the case of a
           period-doubling cascade}
            {Dynamic Bifurcations}
            {\author{E.}{Beno\^\i t}}
            {Springer}{Berlin}{pp107--130}

%B106
\refartbook{refB106}
{\author{E}{Beno\^\i t)}}{1991}
            {Linear dynamic bifurcation with noise}
            {Dynamic Bifurcations}
            {\author{E.}{Beno\^\i t}}
            {Springer}{Berlin}{pp131--150}

\refbook{refB107}
{\author{Denis R.}{Bell}}{1987}
            {The Malliavin Calculus}
            {Longman}{New York}
 
\refarticle{refB108}
{\author{A.R.}{Bulsara} \author{W.C.}{Schieve} \and
\author{E.W.}{Jacobs}}{1990}
            {Homoclinic chaos in a system perturbed by weak Langevin noise}
            {Phys. Rev. A}{41}{668--681}

\refarticle{refB109}
{\author{Per}{Bak}, \author{Chao}{Tang} \and
\author{Kurt}{Wiesenfeld}}{1988}
            {Self-organized criticality}
            {Phys. Rev. A}{38}{364--374}

\refbook{refB110}
{\author{Nicola}{Bellomo} \and \author{Rioccardo}{Riganti}}{1987}
            {Nonlinear Stochastic Systems in Physics and Mathematics}
            {World Scientific}{Singapore}
 
\refarticle{refB111}
{\author{S.M.}{Baer}, \author{T.}{Erneux}}{1986}
            {Singular Hopf bifurcation to relaxation oscillations}
            {SIAM J. Appl. Math.}{46}{721--739}

\refarticle{refB112}
{\author{S.M.}{Baer}, \author{T.}{Erneux}}{1992}
            {Singular Hopf bifurcation to relaxation oscillations II}
            {SIAM J. Appl. Math.}{52}{1651--1664}

%B113
\refarticle{refB113}
{\author{Claude}{Baesens}}{1991}
            {Slow sweep through a period-doubling cascade:
Delayed bifurcations and renormalisation}
            {Physica D}{53}{319--375}

%B114
\refarticle{refB114}
{\author{Andreas}{Becker} \and \author{Lorenz}{Kramer}}{1994}
            {Linear stability analysis for bifurcations in systems
with fluctuating control parameter}
            {Phys. Rev. Lett.}{73}{955--958}

%B115
\refarticle{refB115}
{\author{Paul}{Bryant} \and \author{Kurt}{Wiesenfeld}}{1986}
            {Suppression of period-doubling and nonlinear parametric
effects in periodically perturbed systems}
            {Phys. Rev. A}{33}{2525--2543}

\refbook{refB200}
{\author{E.}{Beno\^\i t (Ed.)}}{1991}
            {Dynamic bifurcations}
            {Springer}{Berlin}
 
%C23
\refbook{refC23}
{\author{J.}{Carr}}{1981}
    {Applications of Centre Manifold Theory}
    {Springer}{New York}
 
%C30
\refphd{refC30}
{\author{F.}{Cattaneo}}{1984}
            {Compressible magnetoconvection}
            {University Cambridge}
 
%C31
\refartbook{refC31}
{\author{F.}{Cattaneo}}{1984}
            {Oscillatory convection in sunspots}
            {The Hydromagnetics of the Sun}
            {\author{T.D.}{Guyenne} \and \author{J.J.}{Hunt}}
            {ESA}{Noordwijkerhoute}{47--50}
 
%C22
\refbook{refC22}
{\author{S.}{Chandrasekhar}}{1961}
            {Hydrodynamic and Hydromagnetic Stability}
            {Clarendon Press}{Oxford}
 
%C17
\refarticle{refC17}
{\author{J.}{Coste} \and \author{N.}{Peyraud}}{1982}
            {A new type of period-doubling bifurcations in one-dimensional
             transformations with two extrema}
            {Physica}{5D}{415--420}
 
%C3
\refarticle{refC3}
{\author{P.H.}{Coullet} \and \author{E.A.}{Spiegel}}{1983}
            {Amplitude equations for systems with competing instabilities}
            {SIAM J.~Appl. Math.}{43}{776--821}
 
%C15
\refarticle{refC15}
{\author{P.}{Coullet}, \author{J.-M.}{Gambaudo} \and \author{C.}{Tresser}}{1984}
            {Une nouvelle bifurcation de codimension~2: le collage de cycles}
            {C.~R.~Acad. Sc. Paris}{299}{S\'erie~I, 253--256}
 
%C7
\refarticle{refC7}
{\author{J.H.}{Curry}, \author{J.R.}{Herring},
 \author{J.}{Loncaric} \and \author{S.A.}{Orszag}}{1984}
            {Order and disorder in two- and three-dimensional B\'enard
             convection}
            {J.~Fluid Mech.}{147}{1--38}

%C100
\refbook{refC100}
{\author{P.}{Collet} \and \author{J.-P.}{Eckmann}}{1986}
    {Iterated maps of the interval as dynamical systems}
    {Birkh\ddot auser}{Boston}
 
%C101
\refarticle{refC101}
{\author{J.P.}{Crutchfield}, \author{J.D.}{Farmer}}{1984}
            {Fluctuations and simple chaotic dynamics}
            {Phys. Rep.}{92}{45--82}
%C102
\refarticle{refC102}
{\author{Alex}{Craik}}{1992}
            {Second-harmonic resonance in non-conservative systems}
            {Wave Motion}{15}{173--183}

%C103
\refphd{refC103}
{\author{Carston C.}{Chow}}{1985}
            {Spatiotemporal chaos on the nonlinear three wave interaction}
            {Massachusetts Institute of Technology}

%C104
\refbook{refC104}
{\author{C.}{Canuto et al}}{1988}
    {Spectral Methods in Fluid Dynamics}
    {Springer}{New York}

%C105
\refarticle{refC105}
{\author{Hugues}{Chat\'e} and \author{Paul}{Manneville}}{1987}
            {Transition to Turbulence via Spatiotemporal Intermittency}
            {Phys. Rev. Lett.}{58}{112--115}

%C106
\refarticle{refC106}
{\author{M.C.}{Cross} \and \author{P.C.}{Hohenberg}}{1993}
            {Pattern formation outside of equilibrium}
            {Rev. Mod. Phys.}{92}{851--1089}

%C107
\refarticle{refC107}
{\author{M.}{Ciofini}, \author{R.}{Meucci}
\and \author{F.T.}{Arecchi}}{1990}
            {Transient statistics in a $CO_2$ laser with a slowly
swept pump}
            {Phys. Rev. A}{42}{482--486}

%C108
\refarticle{refC108}
{\author{Jack}{Carr} \and \author{Robert}{Pego}}{1992}
            {Self-similarity in a coarsening model in one dimension}
            {Proc. Royal Soc. London A}{436}{569--582}

%C109
\refarticle{refC109}
{\author{J.}{Carr} \and \author{R.L.}{Pego}}{1989}
            {Metastable Patterns in Solutions of $u_t=\ep^2u_{xx}-f(u)$}
            {Commun. Pure Appl. Math.}{42}{523--576}

%C111
\refarticle{refC111}
{\author{Jack}{Carr} \and \author{Robert}{Pego}}{1992}
            {Invariant manifolds for metastable patterns in
        $u_t=\ep^2u_{xx}-f(u)$}
            {Proc. Royal Soc. Edinburgh A}{116}{133--160}
%C110
\refbook{refC110}
{\author{H.S.}{Carslaw} \and \author{J.C.}{Jaeger}}{1959}
    {Conduction of heat in solids}
    {Oxford University Press}{London}

%C111
\refarticle{refC111}
{\author{J.P.}{Crutchfield}, \author{B.A.}{Huberman}}{1980}
            {Fluctuations and the onset of chaos}
            {Phys. Lett. A}{77}{407--410}

%C112
\refarticle{refC112}
{\author{Thomas}{Clune}, \author{Edgar}{Knobloch}}{1994}
            {pattern selection in three-dimensional magnetoconvection}
            {Physica D}{74}{151--176}

%D2
\refarticle{refD2}
{\author{L.N.}{Da~Costa}, \author{E.}{Knobloch} \and \author{N.O.}{Weiss}}{1981}
            {Oscillations in double-diffusive convection}
            {J.~Fluid Mech.}{109}{25--43}
 
%D14
\refarticle{refD14}
{\author{G.}{Dangelmayr}, \author{D.}{Armbruster} \and
 \author{M.}{Neveling}}{1985}
            {A codimension three bifurcation for the laser with saturable
             absorber}
            {Z.~Phys.~B -- Condensed Matter}{59}{365--370}
 
%D3
\refarticle{refD3}
{\author{G.}{Dangelmayr} \and \author{E.}{Knobloch}}{1986}
            {Interaction between standing and travelling waves and steady
             states in magnetoconvection}
            {Phys. Lett.}{117A}{394--398}
 
%D1
\refarticle{refD1}
{\author{G.}{Dangelmayr} \and \author{E.}{Knobloch}}{1987}
            {The Takens--Bogdanov bifurcation with $O(2)$-symmetry}
            {Phil. Trans. R.~Soc. Lond.}{A322}{243--279}
 
%D6
\refarticle{refD6}
{\author{A.E.}{Deane}, \author{E.}{Knobloch} \and \author{J.}{Toomre}}{1988}
            {Travelling waves in large-aspect-ratio thermosolutal convection}
            {Phys. Rev.}{37A}{1817--1820}
 
%D11
\refarticle{refD11}
{\author{A.E.}{Deane} \and \author{L.}{Sirovich}}{1991}
            {A computational study of Rayleigh--B\'enard convection. Part~1.
             Rayleigh-number scaling}
            {J.~Fluid Mech.}{222}{231--250}
 
%D15
\refarticle{refD15}
{\author{P.}{Deuflhard}}{1983}
            {Order and stepsize control in extrapolation methods}
            {Numer. Math.}{41}{399--422}
 
%D13
\refbook{refD13}
{\author{E.}{Doedel}}{1986}%
            {AUTO: Software for Continuation and Bifurcation Problems in
             Ordinary Differential Equations}
            {C.I.T. Press}{Pasadena}

\refarticle{refD100}
{\author{M.I.}{Dykman},
\author{R.}{Mannella}, \author{P.V.E.}{McClintock},
\author{N.D.}{Stein} \and \author{N.G.}{Stocks}}{1993}
           {Probability distributions and escape rates for systems
driven by quasimonochromatic noise}
           {Phys. Rev. E}{47}{3996--4009}

\refarticle{refD101}
{\author{I.T.}{Drummond}}{1993}
           {Multiplicative stochastic differential equations with
noise-induced transitions}
           {J. Phys. A}{25}{2273--2296}

\refarticle{refD102}
{\author{S.M.}{Catterall}, \author{I.T.}{Drummond}
\and \author{R.R.}{Horgan}}{1991}
           {Stochastic simulation of quantum mechanics}
           {J. Phys. A}{24}{4081--4091}

\refarticle{refD103}
{\author{Didier}{Dangoisse}, \author{Pierre}{Glorieux}
\and \author{Daniel}{Hennequin}}{1987}
           {Chaos in a $CO_2$ laser with modulated control parameters:
        Experiments and numerical simulations}
           {Phys. Rev. A}{36}{4775--4791}

\refarticle{refD104}
{\author{Charles R.}{Doering}}{1987}
           {Nonlinear Parabolic Stochastic Differential Equations
with Additive Colored Noise on $\real^d\times\real_+$:
A Regulated Stochastic Quantization}
           {Commun. Math. Phys.}{109}{537--561}

\refarticle{refD105}
{\author{Charles R.}{Doering}}{1987}
           {A stochastic partial differential equation
with multiplicative noise}
           {Phys. Lett. A}{122}{133--139}

\refarticle{refD106}
{\author{Marc}{Diener}}{1984}
           {The Canard Unchained or How Fast/Slow Dynamical Systems Bifurcate}
           {The Mathematical Intelligencer}{6}{No.2, 38--49}

\refartbook{refD107}
{\author{C.R.}{Doering}}{1992}
            {Evaluating the Path Integral:
Field Theory Estimates via Stochastic Quantization}
            {Lectures on Path Integration}
            {\author{R.}{Cerdeira}}
            {World Scientific}{Singapore}{pp184--214}

\refartbook{refD108}
{\author{Marc}{Diener}}{1981}
            {On the Perfect Delay Convention, or
the Revolt of the Slaved Variables}
            {Chaos and Order in Nature}
            {\author{H.}{Haken}}
    {Springer}{Berlin}{pp249--268}

%D109
\refartbook{refD109}
{\author{B.}{Candelpergher}, \author{F.}{Diener} \and
\author{M.}{Diener}}{1990}
{Retard \`a la bifurcation: du local au global}
            {Bifurcations of planar vector fields}
            {\author{J.P.}{Francoise} \and \author{R.}{Roussarie}}
            {Springer}{Berlin}{pp1--19}

%D110
\refarticle{refD110}
{\author{E.}{Beno\^it}, \author{F.}{Diener} \and
\author{M.}{Diener}}{1981}
{Chasse au canard}
            {Collectanea Mathematica}
            {31}{37--119}

%D111
\refbook{refD111}
{\author{P.}{Drazin}}{1992}
    {Nonlinear Systems}
    {Cambridge University Press}{Cambridge}

%E100
\refarticle{refE100}
{\author{A.}{Einstein}}{1905}
           {\"uber die von der molekular-kinetischen Theorie der
        W\"arme geforderte Bewegung von in ruhended Fl\"ussigkeiten}
           {Ann.Phys.}{17}{549-555}

%E101
\refartbook{refE101}
{\author{T.}{Erneux}, \author{E.C.}{Reiss}, \author{L.J.}{Holden}
\and \author{M.}{Georgiou}}{1991}
            {Slow Passage through bifurcation and limit points.
Asymptotic theory}
            {Dynamic Bifurcations}
            {\author{E.}{Beno\^\i t}}
            {Springer}{Berlin}{pp14--28}
 
\refarticletobe{refE102}
{\author{Stephen}{Einchcomb} \and \author{A.J.}{McKane}}{1993}
           {Quasi-monochromatic noise stuff with path integrals}
           {Phys. Rev. E}%{48?}{**}

%E103
\refartbook{refE103}
{\author{W.}{Eckhaus}}{1983}
{Relaxation oscillations including a standard chase on french ducks}
            {Asymptotic Analysis}
            {\author{F.}{Verhulst}}
            {Springer}{Berlin}{pp432--449}

%E104
\refarticle{refE104}
{\author{T.}{Erneux} \and \author{Paul}{Mandel}}{1984}
            {Slow passage through laser first threshold}
            {Phys. Rev. A}{39}{5179--5188}

%F5
\refarticle{refF5}
{\author{A.C.}{Fowler}, \author{J.D.}{Gibbon} \and
 \author{M.J.}{McGuinness}}{1983}
            {The real and complex Lorenz equations and their relevance to
             physical systems}
            {Physica}{7D}{126--134}
 
%F4
\refarticle{refF4}
{\author{A.C.}{Fowler}}{1990}
            {Homoclinic bifurcations in $n$ dimensions}
            {Studies in Applied Mathematics}{83}{193--209}

%F100
\refarticle{refF100}
{\author{L.}{Fronzoni}, \author{Frank}{Moss} \and
\author{P.V.E.}{McClintock}}{1987}
{Swept-parameter-induced postponements and noise on the Hopf bifurcation}
            {Phys. Rev. A}{36}{1492--1494}

%F101
\refarticle{refF101}
{\author{K.}{Fujimura} \and \author{R.E.}{Kelly}}{1994}
{Interaction between longitudinal convection rolls and transverse
waves in unstably stratified plane Poiseuille flow}
            {Phys. Fluids}{7}{25--65}

%F102
\refarticle{refF102}
{\author{Ronald Forrest}{Fox}}{1972}
{Contributions to the Theory of Multiplicative Stochastic Processes}
            {J. Math. Phys}{13}{1196--1207}
 
%F103
\refarticle{refF103}
{\author{Tadahisa}{Funaki}}{1983}
{Random motion of strings and related stochastic evolution equations}
            {Nagoya Math. J.}{89}{129--193}

%F104
\refarticle{refF104}
{\author{Tadahisa}{Funaki}}{1995}
{The scaling limit for a stochastic PDE and the separation of phases}
            {Probab. Theory Relat. Fields}{102}{221--288}

%G32
\refarticle{refG32}
{\author{P.}{Gaspard}, \author{R.}{Kapral} \and \author{G.}{Nicolis}}{1984}
            {Bifurcation phenomena near homoclinic systems: a two-parameter
             analysis}
            {J.~Stat. Phys.}{35}{697--727}

%G12
\refarticle{refG12}
{\author{P.}{Glendinning} \and \author{C.}{Sparrow}}{1984}
            {Local and global behaviour near homoclinic orbits}
            {J.~Stat. Phys.}{35}{645--696}
 
%G22
\refphd{refG22}
{\author{P.A.}{Glendinning}}{1985}
            {Homoclinic bifurcations}
            {University of Cambridge}
 
%G2
\refarticle{refG2}
{\author{P.}{Glendinning} \and \author{C.}{Sparrow}}{1986}
            {T-points: a codimension two heteroclinic bifurcation}
            {J.~Stat. Phys.}{43}{479--488}
 
%G45
\refarticle{refG45}
{\author{I.}{Goldhirsch}, \author{R.B.}{Pelz}, \author{S.A.}{Orszag}}{1989}
            {Numerical simulation of thermal convection in a two-dimensional
             finite box}
            {J.~Fluid Mech.}{199}{1--28}
 
%G31
\refbook{refG31}
{\author{D.}{Gottlieb} \and \author{S.A.}{Orszag}}{1977}
    {Numerical Analysis of Spectral Methods: Theory and Applications}
    {Society for Industrial and Applied Mathematics}{Philadelphia}
 
%G41
\refarticle{refG41}
{\author{E.}{Graham}}{1975}
            {Numerical simulation of two-dimensional compressible convection}
            {J.~Fluid Mech.}{70}{689--703}
 
%G42
\refarticle{refG42}
{\author{R.}{Graham}}{1976}
            {Onset of self-pulsing in lasers and the Lorenz model}
            {Phys. Lett.}{58A}{440--442}
 
%G43
\refarticle{refG43}
{\author{S.}{De~Gregorio}, \author{E.}{Scoppola} \and
 \author{B.}{Tirozzi}}{1983}
            {A rigorous study of periodic orbits by means of a computer}
            {J.~Stat. Phys.}{32}{25--33}
 
%G28
\refartbook{refG28}
{\author{J.}{Guckenheimer}}{1976}
            {A strange, strange attractor}
            {The Hopf Bifurcation and Its Applications}
            {\author{J.E.}{Marsden} \and \author{M.}{McCracken}}
            {Springer}{New York}{368--381}
 
%G30
\refarticle{refG30}
{\author{J.}{Guckenheimer} \and \author{E.}{Knobloch}}{1983}
            {Nonlinear convection in a rotating layer: amplitude expansions and
             normal forms}
            {Geophys. Astrophys. Fluid Dynamics}{23}{247--272}
 
%G1
\refbook{refG1}
{\author{J.}{Guckenheimer} \and \author{P.}{Holmes}}{1987}
            {Nonlinear Oscillations,
             Dynamical Systems and Bifurcations of Vector Fields}
            {Springer}{New York}

%G100
\refbook{refG100}
{\author{C.W.}{Gardiner}}{1990}
    {Handbook of Stochastic Methods}
    {Springer}{Berlin}

%G101
\refbook{refG101}
{\author{I.S.}{Gradshteyn} \and \author{M.}{Ryzhik}}{1980}
            {Table of integrals, series, and products} 
            {Academic Press}{New York}

\refartbook{refG102}
{\author{Paolo}{Grigolini}}{1989}
            {The projection approach to the Fokker-Planck equation:
application to phenomenological stochastic equations with colored noise}
            {Noise in nonlinear dynamical systems, Vol. I}
            {\author{Frank}{Moss} \and \author{P.V.E.}{McClintock}}
            {Cambridge University Press}{Cambridge}{pp161--190}

\refartbook{refG103}
{\author{Robert}{Graham}}{1989}
            {Macroscopic potentials, bifurcations and noise in
dissipative systems}
            {Noise in nonlinear dynamical systems, Vol. I}
            {\author{Frank}{Moss} \and \author{P.V.E.}{McClintock}}
            {Cambridge University Press}{Cambridge}{225--278}
  
%G104
\refarticle{refG104}
{\author{I.}{Gy\"ongy} \and \author{E.}{Pardoux}}{1993}
            {On quasi-linear stochastic partial differential equations}
            {Probab. Theory Relat. Fields}{94}{413--426}

%G105
\refarticle{refG105}
{\author{I.}{Gy\"ongy} \and \author{E.}{Pardoux}}{1993}
            {On the regularization effect of space-time white noise on
quasi-linear parabolic partial differential equations}
            {Probab. Theory Relat. Fields}{97}{211--230}

%G106
\refarticle{refG106}
{\author{R.}{Graham} \and \author{T.}{T\'el}}{1990}
            {Steady-state ensemble for the complex Ginzburg-Landau
equation with weak noise}
            {Phys. Rev. A}{42}{4661--4677}

%G107
\refarticle{refG107}
{\author{Hermann}{Grabert} \author{Peter}{H\ddot anggi}
\and \author{Peter}{Talkner}}{1979}
           {Is quantum mechanics equivalent to a classical stochastic process?}
            {Phys. Rev. A}{19}{2440--2445}

%G108
\refarticle{refG108}
{\author{J.}{Garc\'\i a-Ojalvo}
\and \author{J.M.}{Sancho}}{1994}
           {Colored noise in spatially extended systems}
            {Phys. Rev. E}{49}{2769--2778}

%G109
\refarticle{refG109}
{\author{J.}{Garc\'\i a-Ojalvo}, \author{J.M.}{Sancho}
\and \author{L.}{Ram\'\i rez-Piscina}}{1992}
           {Generation of spatiotemporal colored noise}
            {Phys. Rev. A}{46}{4670--4675}

%G110
\refarticle{refG110}
{\author{Robert}{Graham}}{1974}
            {Hydrodynamic fluctuations near the convection instability}
            {Phys. Rev. A}{10}{1762--1784}
%G111
\refarticle{refG111}
{\author{R.}{Graham} \and \author{A.}{Schenzle}}{1982}
            {Stabilization by multiplicative noise}
            {Phys. Rev. A}{26}{1676--1685}

%G112
\refarticle{refG112}
{\author{J.}{Garc\'\i a-Ojalvo}, \author{A.}{Hern\'andez-Machado} \and
\author{J.M.}{Sancho}}{1993} 
           {Effects of External Noise on the Swift-Hohenberg equation}
            {Phys. Rev. Lett.}{71}{1542--1545}

%G113
\refarticle{refG113}
{\author{P.}{Gaspard} et al (Ed.)}{1991}
            {Homoclinic Chaos}
            {Physica D}{62}{1--372}

%G114
\refarticle{refG114}
{\author{Raymind}{Goldstein} \and
\author{Gemunu H.}{Gunaratne}}{1991} 
           {Hydrodynamic and interfacial patterns with broken
space-time symmetry}
            {Phys. Rev. A}{43}{6700--6710}

%H13
\refarticle{refH13}
{\author{H.}{Haken}}{1975}
            {Analogy between higher instabilities in fluids and lasers}
            {Phys . Lett.}{53A}{77--78}
 
%H17
\refarticle{refH17}
{\author{S.M.}{Hammel}, \author{J.A.}{Yorke} \and \author{C.}{Grebogi}}{1988}
            {Numerical orbits of chaotic processes represent true orbits}
            {Bull. Am. Math. Soc.}{19}{465--469}
 
%H15
\refbook{refH15}
{\author{A.C.}{Hearne}}{1987}
            {Reduce User's Manual -- Version 3.3}
            {Rand}{Santa Monica}
 
%H3
\refarticle{refH3}
{\author{D.W.}{Hughes} \and \author{M.R.E.}{Proctor}}{1988}
            {Magnetic fields in the solar convection zone: magnetoconvection
             and magnetic buoyancy}
            {Ann. Rev. Fluid Mech.}{20}{187--223}
 
%H5
\refarticle{refH5}
{\author{H.E.}{Huppert} \and \author{D.R.}{Moore}}{1976}
            {Nonlinear double-diffusive convection}
            {J.~Fluid Mech.}{78}{821--854}
 
%H16
\refarticle{refH16}
{\author{N.E.}{Hurlburt} \and \author{J.}{Toomre}}{1988}
            {Magnetic fields interacting with nonlinear compressible convection}
            {Ap. J.}{327}{920--932}
 
%H1
\refarticle{refH1}
{\author{N.E.}{Hurlburt}, \author{M.R.E.}{Proctor},
 \author{N.O.}{Weiss} \and \author{D.P.}{Brownjohn}}{1989}
            {Nonlinear compressible magnetoconvection.
             Part~1. Travelling waves and oscillations}
            {J.~Fluid Mech.}{207}{587--628}

%H100
\refbook{refH100}
{\author{H.}{Haken(ed)}}{1981}
    {Chaos and Order in Nature}
    {Springer}{Berlin}
 
%H101
\refarticle{refH101}
{\author{D.W.}{Hughes} \and \author{M.R.E.}{Proctor}}{1990}
            {Chaos and the effect of noise in a model of three-wave
mode coupling}
            {Physica D}{46}{163--176}

%H102
\refarticle{refH102}
{\author{D.W.}{Hughes} \and \author{M.R.E.}{Proctor}}{1990}
            {A low order model of the shear instability of convection:
chaos and the effect of noise}
            {Nonlinearity}{3}{127--153}

%H103
\refarticle{refH103}
{\author{M.R.E.}{Proctor} \and\author{D.W.}{Hughes}}{1990}%
{The false Hopf bifurcation and noise sensitivity
        in bifurcations with symmetry}
            {Eur.J.Mech.,B/Fluids}{10}{81--86}

%H104
\refarticle{refH104}
{\author{Rebecca L.}{Honeycutt}}{1992}
            {Stochastic Runge-Kutta algorithms. I. White noise}
            {Phys. Rev. A}{45}{600--603}

%H105
\refbook{refH105}
{\author{Harry}{Hochstadt}}{1971}
    {The functions of mathematical physics}
    {Dover}{New York}

\refartbook{refH106}
{\author{Peter}{Hanggi}}{1989}
            {Colored noise in continuous dynamical systems: a
functional calculus approach}
            {Noise in nonlinear dynamical systems, Vol. I}
            {\author{Frank}{Moss} \and \author{P.V.E.}{McClintock}}
            {Cambridge University Press}{Cambridge}{307--328}
 
%H107
\refbook{refH107}
{\author{H.}{Haken}}{1978}
    {Synergetics}
    {Springer}{Berlin}

%H108
\refarticle{refH108}
{\author{Richard}{Haberman}}{1979}
            {Slowly varying jump and transition phenomena associated
with algebraic bifurcation problems}
            {SIAM J. Appl. Math.}{37}{69--106}

%H109
\refarticle{refH109}
{\author{J.J.}{Healy}, \author{D.S.}{Broomhead}, \author{K.A.}{Cliff},
\author{R.}{Jones} \and \author{T.}{Mullin}}{1991}
            {The origins of chaos in a modified Van der Pol oscillator}
            {Physica D}{48}{322--339}

%H110
\refbook{refH110}
{\author{W.}{Horsthemke} \& \author{R.}{Lefever}}{1984}
    {Noise-induced transitions}
    {Springer}{Berlin}
 
%H111
\refarticle{refH111}
{\author{W.}{Horsthemke} \and \author{R.}{Lefever}}{1977}
    {Phase transitions induced by external noise}
        {Phys. Lett. A}{64}{19--21}
 
%H112
\refarticle{refH112}
{\author{Michael R. E.}{Proctor} \and \author{Judith Y.}{Holyer}}{1986}
    {Planform selection in salt fingers}
        {J. Fluid Mech.}{168}{241--253}
%H113
\refarticle{refH113}
{\author{Lisa}{Holden} \and \author{Thomas}{Erneux}}{1993}
    {Slow passage through a Hopf bifurcation:
from oscillatory to steady state solutions}
        {SIAM J. Appl. Math.}{53}{1045--1058}
 
%H114
\refarticle{refH114}
{\author{P.}{Holmes}}{1979}
    {A nonlinear oscillator with a strange attractor}
        {Phil. Trans.}{292}{419--448}
%H115
\refarticle{refH115}
{\author{P.}{Holmes}}{1990}
    {Poincar\'e, celestial mechanics, dynamical-systems theory and `chaos'}
        {Phys. Rep.}{193}{137--163}
 
%H116
\refarticle{refH116}
{\author{L.N.}{Howard} \and \author{R.}{Krishnamurti}}{1986}
    {Large-scale flow in turbulent convection:
a mathematical model}
        {J. Fluid Mech.}{170}{385--410}

%H117
\refarticle{refH117}
{\author{P.C.}{Hohenberg} \and \author{J.W.}{Swift}}{1992}
            {Effects of additive noise at the onset of 
Rayleigh-B\'enard convection}
            {Phys. Rev. A}{46}{4773--4785}

%I100
\refarticle{refI100}
{\author{Kiyosi}{Ito}}{1964}
            {Expected number of zeros of continuous stationary Gaussian processes}
            {J. Math. Kyoto Univ.}{3-2}{207--216}
 
%J100
\refarticle{refJ100}
{\author{Peter}{Jung}, \author{George}{Gray}, \author{Rajarshi}{Roy}
 \and \author{Paul}{Mandel}}{1990}
    {Scaling Law for Dynamical Hysteresis}
        {Phys. Rev. Lett.}{65}{1873--1876}

%J5
\refphd{refJ5}
{\author{K.A.}{Julien}}{1991}
            {Strong spatial resonance in convection}
            {University of Cambridge}
 
%K2
\refarticle{refK2}
{\author{E.}{Knobloch} \and \author{M.R.E.}{Proctor}}{1981}
            {Nonlinear periodic convection in double-diffusive systems}
            {J.~Fluid Mech.}{108}{291--316}
 
%K5
\refarticle{refK5}
{\author{E.}{Knobloch}, \author{N.O.}{Weiss} \and \author{L.N.}{Da~Costa}}{1981}
            {Oscillatory and steady convection in a magnetic field}
            {J.~Fluid Mech.}{113}{153--186}
 
%K7
\refarticle{refK7}
{\author{E.}{Knobloch} \and \author{N.O.}{Weiss}}{1983}
            {Bifurcations in a model of magnetoconvection}
            {Physica}{9D}{379--407}
 
%K11
\refarticle{refK11}
{\author{E.}{Knobloch} \and \author{N.O.}{Weiss}}{1984}
            {Convection in sunspots and the origin of umbral dots}
            {Mon.~Not. R.~Astr. Soc.}{207}{203--214}
 
%K8
\refarticle{refK8}
{\author{E.}{Knobloch}}{1986}
            {On convection in a horizontal magnetic field with periodic
             boundary conditions}
            {Geophys. Astrophys. Fluid Dynamics}{36}{161--177}
 
%K19
\refarticle{refK19}
{\author{E.}{Knobloch}}{1986}
            {On the degenerate Hopf bifurcation with $O(2)$ symmetry}
            {Contemp. Math.}{56}{193--201}
 
%K24
\refarticle{refK24}
{\author{E.}{Knobloch}}{1986}
            {Normal forms for bifurcations at a double zero eigenvalue}
            {Phys. Lett.}{115A}{199--201}
 
%K8a
\refarticle{refK8a}
{\author{E.}{Knobloch}}{1986a}
            {On convection in a horizontal magnetic field with periodic
             boundary conditions}
            {Geophys. Astrophys. Fluid Dynamics}{36}{161--177}
 
%K19b
\refarticle{refK19b}
{\author{E.}{Knobloch}}{1986b}
            {On the degenerate Hopf bifurcation with $O(2)$ symmetry}
            {Contemp. Math.}{56}{193--201}
 
%K24c
\refarticle{refK24c}
{\author{E.}{Knobloch}}{1986c}
            {Normal forms for bifurcations at a double zero eigenvalue}
            {Phys. Lett.}{115A}{199--201}
 
%K4
\refarticle{refK4}
{\author{E.}{Knobloch}, \author{D.R.}{Moore},
 \author{J.}{Toomre} \and \author{N.O.}{Weiss}}{1986}
            {Transitions to chaos in two-dimensional double-diffusive
             convection}
            {J.~Fluid Mech.}{166}{409--448}
 
%K1
\refarticle{refK1}
{\author{E.}{Knobloch} \and \author{M.R.E.}{Proctor}}{1988}
            {The double Hopf bifurcation with $2:1$ resonance}
            {Proc. R.~Soc. Lond.}{415A}{61--90}
 
%K17
\refarticle{refK17}
{\author{E.}{Knobloch} \and \author{M.}{Silber}}{1990}
            {Travelling wave convection in a rotating layer}
            {Geophys. Astrophys. Fluid Dynamics}{51}{195--209}

%K34
\refbook{refK34}
{\author{D.E.}{Knuth}}{1984}
            {The \TeX book}
            {Addison Wesley}{Reading}
%K100
\refbook{refK100}
{\author{Peter E.}{Kloeden} \and \author{Eckhard}{Platen}}{1992}
    {Numerical Solution of Stochastic Differential Equations}
    {Springer}{Berlin}

%K101
\refbook{refK101}
{\author{H.}{Kunita}}{1990}
    {Stochastic Flows and Stochastic Differential Equations}
    {Cambridge University Press}{Cambridge}

%K102
\refbook{refK102}
{\author{T.W.}{K\ddot orner}}{1988}
    {Fourier analysis}
    {Cambridge University Press}{Cambridge}

%K103
\refarticle{refK103}
{\author{N.}{van Kampen}}{1981}
            {\ito\ Versus Stratonovich}
            {J. Stat. Phys.}{24}{175--187}
%K104
\refarticle{refK104}
{\author{D.K.}{Kondepudi} \and \author{G.W.}{Nelson}}{1985}
            {Weak neutral currents and the origin of biomolecular chirality}
            {Nature}{314}{438--441}
%K105
\refarticle{refK105}
{\author{D.K.}{Kondepudi}, \author{F.}{Moss} \and 
\author{P.V.E.}{McClintock}}{1986}
            {Observation of symmetry breaking, state selection and
sensitivity in a noisy electronic system}
            {Physica D}{21}{296--306}
 
%K106
\refbook{refK106}
{\author{Frank B.}{Knight}}{1981}
    {Essentials of Brownian motion and diffusion}
    {American Mathematical Society}{Providence}

%K107
\refarticle{refK107}
{\author{Raymond}{Kapral} \and \author{Paul}{Mandel}}{1985}
            {Bifurcation structure of the nonautonomous quadratic map}
            {Phys. Rev. A}{32}{1076--1081}

%K108
\refbook{refK108}
{\author{J.}{Kevorkian} \and \author{J.D.}{Cole}}{1981}
    {Perturbation Methods In Applied Mathematics}
    {Springer}{New York}

%K109
\refarticle{refK109}
{\author{D.J.}{Kaup}, \author{A.}{Rieman} \and \author{A.}{Bers}}{1979}
            {Space-time evolution of nonlinear three-wave interactions}
            {Rev. Mod. Phys.}{51}{275--309}
 
%K110
\refbook{refK110}
{\author{Ioannis}{Karatzas} \and \author{Steven E.}{Shreve}}{1988}
    {Brownian Motion and Stochastic Calculus}
    {Springer}{New York}

%L5
\refarticle{refL5}
{\author{E.N.}{Lorenz}}{1963}
            {Deterministic nonperiodic flow}
            {J.~Atmos. Sci.}{20}{130--141}
 
%L12
\refarticle{refL12}
{\author{D.V.}{Lyubimov} \and \author{M.A.}{Zaks}}{1983}
            {Two mechanisms of the transition to chaos in finite-dimensional
             models of convection}
            {Physica}{9D}{52--64}
 
%L18
\refarticle{refL18}
{\author{D.V.}{Lyubimov}, \author{A.S.}{Pikovsky} \and
 \author{M.A.}{Zaks}}{1989}
            {Universal scenarios of transition to chaos via homoclinic
             bifurcations}
            {Sov. Sci. Rev.~C. Math. Phys.}{8}{221--292}

%L23
\refarticle{refL23}
{\author{A.S.}{Landsberg} \and \author{E.}{Knobloch}}{1991}
            {Direction-reversing traveling waves}
            {Phys. Lett. A}{159}{17--20}

%L24
\refarticle{refL24}
{\author{A.S.}{Landsberg} \and \author{E.}{Knobloch}}{1993}
            {New types of waves in systems with $O(2)$ symmetry}
            {Phys. Lett. A}{179}{316--324}

%L100
\refarticle{refL100}
{\author{G.D.}{Lythe} \and \author{M.R.E.}{Proctor}}{1993}
            {Noise and slow-fast dynamics in a three-wave resonance problem}
            {Phys. Rev. E. }{47}{3122-3127}
 
%L101
\refarticle{refL101}
{\author{M.R.E.}{Proctor} \and \author{G.D.}{Lythe}}{1993}
            {Noise and resonant mode interactions}
            {Ann. New York Acad. Sci. }{706}{42-53}

%L102
\refartbook{refL102}
{\author{Claude}{Lobry}}{1991}
            {Dynamic Bifurcations}
            {Dynamic Bifurcations}
            {\author{E.}{Beno\^\i t}}
            {Springer}{Berlin}{pp1--13}
 
\refartbook{refL103}
{\author{Katya}{Lindenberg}, {Bruce J.}{West} \and {Jaume}{Masoliver}}{1989}
            {First passage time problems for non-Markovian processes}
            {Noise in nonlinear dynamical systems, Vol. I}
            {\author{Frank}{Moss} \and \author{P.V.E.}{McClintock}}
            {Cambridge University Press}{Cambridge}{110--160}
 
%L104
\refbook{refL104}
{\author{L.}{Landau} \and \author{E.}{Lifshitz}}{1959}
    {Statistical Physics}
    {Pergamon}{London}

%L105
\refarticle{refL105}
{\author{N.R.}{Lebovitz} \and \author{R.J.}{Schaar}}{1975}
            {Exchange of stabilities in autonomous systems}
            {Studies in Appl. Math. }{54}{229--260}

%L106
\refarticle{refL106}
{\author{Claude}{Lobry}}{1992}
        {A propos du sens des textes math\'ematiques. Un example:
        la th\'eorie des ``bifurcations dynamiques''}
            {Ann. Inst. Fourier}{42}{327-352}

%L108
%\refartbook{refL108}
%{\author{G.D.}{Lythe}}{1994}
%            {A noise-controlled dynamic bifurcation}
%            {Fluctuations and Order: The New Synthesis}
%               {\author{M.M.}{Millonas}}{Springer}{Berlin}{}
 
%L109
\refartbook{refL109}
{\author{G.D.}{Lythe}}{1994}
            {A noise-controlled dynamic bifurcation}
            {Chaos and Nonlinear Mechanics}
                {\author{T.}{Kapitaniak} and \author{J.}{Brindley}(eds)}
{World Scientific}{Singapore}{147--158}
 
%L110
\refbook{refL110}
{\author{Andrej}{Lasota} \and \author{Michael C.}{Mackey}}{1985}
    {Probabilistic properties of deterministic systems}
    {Cambridge University Press}{Cambridge}

%L111
\refarticle{refL111}
{\author{Pui-Man}{Lam} \and \author{Diola}{Bagayoko}}{1993}
            {Spatiotemporal correlation of coloured noise}
            {Phys. Rev. E}{48}{3267--3270}
%L112
\refarticle{refL112}
{\author{Katya}{Lindenberg}, {Bruce J.}{West} \and {Raoul}{Kopelman}}{1989}
            {Steady-State Segregation in Diffusion-Limited Reactions}
            {Phys. Rev. Lett.}{60}{1777--1780}

%L113
\refbook{refL113}
{\author{Leon}{Lapidus} \and \author{George F.}{Pinder}}{1982}
    {Numerical Solution of Partial Differential Equations in Science and Engineering}
    {Wiley}{New York}

%L114
\refartbook{refL114}
{\author{G.D.}{Lythe}}{1994}
            {Noise and Dynamic transitions}
            {Stochastic Partial Differential Equations}
                {\author{Alison}{Etheridge}}{Cambridge University Press}
                {Cambridge}{}

\refarticle{refL115}
{\author{G.D.}{Lythe}}{1995}
           {Dynamics controlled by additive noise}
           {Nuovo Cimento D}{17}{855--861}

\refphd{refL116}
{\author{G.D.}{Lythe}}{1994}
            {Stochastic slow-fast dynamics}
            {University of Cambridge}

\refarticletobe{refL117}
{\author{C.-H.}{Lu} \and \author{R.M.}{Evan-Iwanowski}}{1994}
           {The Nonstationary Effects on the Logistic Map and the
Softening Duffing Oscillator}
           {}%{48?}{}

\refartbook{refM28}
{\author{B.A.}{Malomed} \and \author{A.A.}{Nepomnyashchy}}{1990}
            {Onset of chaos in the generalized Ginzburg--Landau equation}
            {Nonlinear Evolution of Spatio-Temporal Structures in Dissipative
             Continuous Systems}
            {\author{F.H.}{Busse} \and \author{L.}{Kramer}}
            {Plenum Press}{New York}{419--424}

%M25
\refarticle{refM25}
{\author{W.V.R.}{Malkus} \and \author{G.}{Veronis}}{1958}
            {Finite amplitude cellular convection}
            {J.~Fluid Mech.}{4}{225--260}
 
%M19
\refarticle{refM19}
{\author{P.S.}{Marcus}}{1981}
            {Effects of truncation in modal representations of thermal
             convection}
            {J.~Fluid Mech.}{103}{241--255}
 
%M28
\refartbook{refM28}
{\author{B.A.}{Malomed} \and \author{A.A.}{Nepomnyashchy}}{1990}
            {Onset of chaos in the generalized Ginzburg--Landau equation}
            {Nonlinear Evolution of Spatio-Temporal Structures in Dissipative
             Continuous Systems}
            {\author{F.H.}{Busse} \and \author{L.}{Kramer}}
            {Plenum Press}{New York}{419--424}
 
%M34
\refbook{refM34}
{\author{P.}{Manneville}}{1990}
            {Dissipative Structures and Weak Turbulence}
            {Academic Press}{Boston}
 
%M5
\refarticle{refM5}
{\author{R.M.}{May}}{1976}
            {Simple mathematical models with very complicated dynamics}
            {Nature}{261}{459--467}
 
%M20
\refarticle{refM20}
{\author{D.R.}{Moore} \and \author{N.O.}{Weiss}}{1973}
            {Two-dimensional Rayleigh--B\'enard convection}
            {J.~Fluid Mech.}{58}{289--312}
 
%M22
\refarticle{refM22}
{\author{D.R.}{Moore}, \author{N.O.}{Weiss} \and \author{J.M.}{Wilkins}}{1990}
            {The reliability of numerical experiments:
             transitions to chaos in thermosolutal convection}
            {Nonlinearity}{3}{997--1014}

%M100
\refarticle{refM100}
{\author{Paul}{Mandel} \and \author{T.}{Erneux}}{1984}
            {Laser Lorenz equations with a time-dependent parameter}
            {Phys. Rev. Lett.}{53}{1818--1820}

%M101
\refarticle{refM101}
{\author{Paul}{Mandel} \and \author{Thomas}{Erneux}}{1987}
            {The slow passage through a steady bifurcation:
        delay and memory effects}
            {J. Stat. Phys.}{48}{1059--1070}

\refarticle{refM102}
{\author{R.}{Mannella}, \author{Frank}{Moss}
\and \author{P.V.E.}{McClintock}}{1987}
           {Postponed bifurcations of a ring-laser with a swept parameter
        and additive colored noise}
           {Phys. Rev. A}{35}{2560--2566}
%M103
\refbook{refM103}
{\author{Frank}{Moss} \and \author{P.V.E.}{McClintock}}{1987}
            {Noise and nonlinear dynamical systems}
            {Cambridge University Press}{Cambridge}

%M104
\refarticle{refM104}
{\author{Miltiades}{Georgiou} \and \author{Thomas}{Erneux}}{1992}
          {Pulsating laser oscillations depend on 
        extremely-small-amplitude noise}
           {Phys. Rev. A}{45}{6636--6642}

%M105
\refbook{refM105}
{\author{A}{Atchison}}{19**}
            {The lognormal distribution}
            {Cambridge University Press}{Cambridge}

%M106
\refarticletobe{refM106}
{\author{Michael C.}{Mackey} \and \author{Irina G.}{Nechaeva}}{1993}
            {Noise and stability in differential delay equations}
             {Journal of Dynamics and Differential Equations}

%M107
\refarticle{refM107}
{\author{Christopher W.}{Meyer}, \author{Guenter}{Ahlers}
\and \author{David S.}{Cannell}}{1991}
            {Stochastic influences on pattern formation in
Rayleigh-B\'enard convection: Ramping experiments}
             {Phys. Rev. A}{44}{2514--2537}

%M108
\refbook{refM108}
{\author{Kenneth S.}{Miller}}{1974}
            {Complex Stochastic Processes}
            {Addison-Wesley}{Reading}
%M208
\refbook{refM208}
{\author{Kenneth S.}{Miller}}{1964}
            {Multidimensional Gaussian Distributions}
            {Wiley}{New York}

\refarticle{refM109}
{\author{R.}{Mannella}
\and \author{P.V.E.}{McClintock}}{1990}
           {Noise in nonlinear dynamical systems}
           {Contemporary Physics}{31}{179--194}

\refarticle{refM110}
{\author{P.C.}{Matthews}, \author{M.R.E.}{Proctor},
\author{A.M.}{Rucklidge} \and \author{N.O.}{Weiss}}{1993}
            {Pulsating waves in nonlinear magnetoconvection}
            {Phys. Lett. A}{183}{69--75}
 
\refarticle{refM111}
{\author{Frank}{Moss}, \author{D.K.}{Kondepudi}
 \and \author{P.V.E.}{McClintock}}{1985}
            {Branch selectivity at the bifurcation of a bistable
system with external noise}
            {Phys. Lett. A}{112}{293--296}
 
\refarticle{refM112}
{\author{Brian}{Morris} \and \author{Frank}{Moss}}{1986}
            {Postponed bifurcations of a quadratic map with a swept parameter}
            {Phys. Lett. A}{118}{117--120}
 
\refarticle{refM113}
{\author{C.}{Meunier} \and \author{A.D.}{Verga}}{1988}
            {Noise and Bifurcations}
            {J. Stat. Phys.}{50}{345--375}
 
%M114
\refarticle{refM114}
{\author{Michael C.}{Mackey}, \author{Andr\'e}{Longtin}
\and \author{Andrzej}{Lasota}}{1990}
            {Noise-Induced Global Asymptotic Stability}
             {J. Stat. Phys.}{60}{735--751}

%N2
\refarticle{refN2}
{\author{R.S.}{Mckay} \and \author{C.}{Tresser}}{1987}
            {Some flesh on the skeleton: the bifurcation structure
                of bimodal maps}
            {Physica}{27D}{412--422}

%N3
\refarticle{refN3}
{\author{M.}{Nagata}, \author{M.R.E.}{Proctor} \and
 \author{N.O.}{Weiss}}{1990}
            {Transitions to asymmetry in magnetoconvection}
            {Geophys. Astrophys. Fluid Dynamics}{51}{211--241}
 
%N2
\refarticle{refN2}
{\author{A.C.}{Newell} \and \author{J.A.}{Whitehead}}{1969}
            {Finite bandwidth, finite amplitude convection}
            {J.~Fluid Mech.}{38}{279--303}
 
%N4
\refarticle{refN4}
{\author{J.}{Niederl\"ander}, \author{M.}{L\"ucke} \and
 \author{M.}{Kamps}}{1991}
            {Weakly nonlinear convection: Galerkin model, numerical simulation,
             and amplitude equation}
            {Z.~Phys.~B -- Condensed Matter}{82}{135--141}

\refarticle{refN100}
{\author{A.I.}{Neishtadt}}{1987}
           {Persistence of stability loss for dynamical bifurcations}
           {Differential Equations}{23}{1385--1391}

\refarticle{refN101}
{\author{Bruce}{McNamara} and {Kurt}{Wiesenfeld}}{1989}
           {Theory of stochastic resonance}
           {Phys. Rev. A}{39}{4854--4869}

%N102
\refarticle{refN102}
{\author{Alan C.}{Newell}, \author{Thierry}{Passot}
 \and \author{Joceline}{Lega}}{1993}
            {Order parameter equations for patterns}
            {Ann. Rev. Fluid Mech.}{25}{399--453}

%N103
\refarticle{refN103}
{\author{Ali H.}{Nayfeh} \and \author{Nestor E.}{Sanchez}}{1989}
            {Bifurcations in a forced softening Duffing oscillator}
            {Int. J. Non-Linear Mechanics}{24}{483--497}
 
%N104
\refarticle{refN104}
{\author{S.}{Novak} \and \author{R. G.}{Frehlich}}{1982}
            {Transition to chaos in the Duffing oscillator}
            {Phys. Rev. A}{26}{3660--3663}
 
%O100
\refbook{refO100}
{\author{B.}{\O ksendal}}{1989}
    {Stochastic Differential Equations}
    {Springer}{Berlin}

%P6
\refbook{refP6}
{\author{T.S.}{Parker} \and \author{L.O.}{Chua}}{1989}
            {Practical Numerical Algorithms for Chaotic Systems}
            {Springer}{New York}
 
%P2
\refbook{refP2}
{\author{W.H.}{Press}, \author{B.P.}{Flannery},
 \author{S.A.}{Teukolsky} \and \author{W.T.}{Vetterling}}{1986}
            {Numerical Recipes -- the Art of Scientific Computing}
            {Cambridge University Press}{Cambridge}
 
%P4
\refarticle{refP4}
{\author{M.R.E.}{Proctor} \and \author{C.A.}{Jones}}{1988}
            {The interaction of two spatially resonant patterns in thermal
             convection. Part~1. Exact $1:2$~resonance}
            {J.~Fluid Mech.}{188}{301--335}
 
%P1
\refarticle{refP1}
{\author{M.R.E.}{Proctor} \and \author{N.O.}{Weiss}}{1982}
            {Magnetoconvection}
            {Rep. Prog. Phys.}{45}{1317--1379}
 
%P8
\refarticle{refP8}
{\author{M.R.E.}{Proctor} \and \author{N.O.}{Weiss}}{1990}
            {Normal forms and chaos in thermosolutal convection}
            {Nonlinearity}{3}{619--637}

\refartbook{refP100}
{\author{M.R.E.}{Proctor} \and \author{D.W.}{Hughes}}{1990}
        {Chaos and the effect of noise for the double Hopf
bifurcation with 2:1 resonance}
        {Nonlinear Evolution of Spatio-Temporal Structures in
Dissipative Continuous Systems}
        {F H Busse and L Kramer}
        {Plenum}{New York}{}
%P101
\refarticle{refP101}
{\author{F.}{de Pasquale}, \author{J.M.}{Sancho}, \author{M.}{San Miguel}
 \and \author{P.}{Tartaglia}}{1986}
            {Decay of an unstable state in the presence of
multiplicative noise}
            {Phys. Rev. A}{33}{4360--4366}

%P102
\refarticle{refP102}
{\author{P.}{Piera\'nski}
 \and \author{J.}{Ma\l ecki}}{1987}
            {Noise-sensitive Hysteresis Loops around Period-Doubling
Bifurcation Points}
            {Il Nuovo Cimento}{9}{757--779}
 
%P103
\refbook{refP103}
{\author{Guiseppe}{Da Prato} \and \author{Jerzy}{Zabczyk}}{1992}
    {Stochastic Equations in Infinite Dimensions}
    {Cambridge University Press}{Cambridge}     

%P104
\refbook{refP104}
{\author{Phoolan}{Prasad} \and \author{Renuka}{Ravindran}}{1985}
    {Partial Differential Equations}
    {Wiley Eastern}{New Delhi}  

%P105
\refarticle{refP105}
{\author{Nathan}{Platt}, \author{Stephen M.}{Hammel}
\and \author{James F.}{Heagy}}{1994}
            {Effect of Additive Noise on On-Off Intermittency}
            {Phys. Rev. Lett.}{52}{3498--3501}

%P106
\refbook{refP106}
{\author{Philip}{Protter}}{1990}
    {Stochastic Integration and Differential Equations}
    {Springer}{Berlin}

%R5
\refarticle{refR5}
{Lord~Rayleigh}{1916}
            {On convection currents in a horizontal layer of fluid, when the
             higher temperature is on the under side}
            {Phil.~Mag.}{32}{529--546}
 
%R13
\refarticle{refR13}
{\author{A.J.}{Rodr\'\i guez-Luis}, \author{E.}{Freire}
 \and \author{E.}{Ponce}}{1991}
            {On a codimension 3 bifurcation arising in an autonomous electronic
             circuit}
            {International Series of Numerical Mathematics}{97}{301--306}
 
%R8
\refarticle{refR8}
{\author{H.T.}{Rossby}}{1969}
            {A study of B\'enard convection with and without rotation}
            {J.~Fluid Mech.}{36}{309--335}
 
%R2
\refarticle{refR2}
{\author{A.}{Rucklidge} \and \author{S.}{Zaleski}}{1988}
            {A microcanonical model for interface formation}
            {J.~Stat. Phys.}{51}{299--307}
 
%R9
\refarticle{refR9}
{\author{A.M.}{Rucklidge}}{1992}
            {Chaos in models of double convection}
            {J.~Fluid Mech.}{237}{209--229}

%R15
%\refarticleshort{refR15}
%{\author{A.M.}{Rucklidge}, \author{N.O.}{Weiss},
% \author{D.P.}{Brownjohn} \and \author{M.R.E.}{Proctor}}{1992}
%            {Oscillations and secondary bifurcations in nonlinear
%             magnetoconvection}
%            {Geophys. Astrophys. Fluid Dynamics {\rm (submitted)}}

%R17
%\refarticleshort{refR17}
%{\author{A.M.}{Rucklidge}}{1992}
%            {Chaos in magnetoconvection}
%            {{\rm (in preparation)}}

\refbook{refR101}
{\author{Daniel}{Revuz} \and \author{Marc}{Yor}}{1991}
            {Continuous Martingales and Brownian Motion}
            {Springer}{Berlin}

%R100
\refbook{refR100}
{\author{H.}{Risken}}{1984}
            {The Fokker-Planck Equation}
            {Springer}{Berlin}
%R102
\refarticle{refR102}
{\author{L.A.}{Rubenfeld}}{1979}
            {A model bifurcation problem exhibiting the effects of
slow passage through critical}
            {SIAM J. Appl. Math.}{37}{302--306}
 
%R103
\refphd{refR103}
{\author{A.M.}{Rucklidge}}{1991}
            {Chaos in models of double convection}
            {University of Cambridge}
  
%R105
\refarticle{refR105}
{\author{A.M.}{Rucklidge} \and \author{P.C.}{Matthews}}{1995}
            {Analysis of the shearing instability
 in nonlinear magnetoconvection}
            {Nonlinearity}{9}{311--352}

%S31
\refarticle{refS31}
{\author{T.}{Sauer} \and \author{J.A.}{Yorke}}{1991}
            {Rigorous verification of trajectories for the computer simulation
             of dynamical systems}
            {Nonlinearity}{4}{961--979}
 
%S26
%\refrussianartbook{refS26}
%{\author{A.L.}{Shil'nikov}}{1986}
%            {Bifurcations and chaos in the Marioka--Shimizu system}
%            {Methods of Qualitative Theory of Differential Equations}
%            {Gorky State University}{180--193. Published in English
%  (1991) {\sl Selecta Math. Sov.} {\bf 10} 105--117}
 
%\refrussianartbook{refS27}
%{\author{A.L.}{Shil'nikov}}{1989}
%            {Bifurcations and chaos in the Marioka--Shimizu model: II}
%            {Methods of Qualitative Theory of Differential Equations and Theory
%             of Bifurcations}
%            {Gorky State University}{130--138}

%S13
\refarticle{refS13}
{\author{L.P.}{Shil'nikov}}{1965}
            {A case of the existence of a countable number of periodic motions}
            {Soviet Maths. Dokl.}{6}{163--167}

%S25
\refarticle{refS25}
{\author{T.}{Shimizu} \and \author{N.}{Morioka}}{1978}
            {Chaos and limit cycles in the Lorenz model}
            {Phys. Lett.}{66A}{182--184}
 
%S18
\refarticle{refS18}
{\author{T.}{Shimizu} \and \author{N.}{Morioka}}{1980}
            {On the bifurcation of a symmetric limit cycle to an
             asymmetric one in a simple model}
            {Phys. Lett.}{76A}{201--204}

%S10
\refarticle{refS10}
{\author{T.G.L.}{Shirtcliffe}}{1969}
            {An experimental investigation of thermosolutal convection at
             marginal stability}
            {J.~Fluid Mech.}{35}{677--688}
 
%S19
\refarticle{refS19}
{\author{L.}{Sirovich} \and \author{A.E.}{Deane}}{1991}
            {A computational study of Rayleigh--B\'enard convection. Part~2.
             Dimension considerations}
            {J.~Fluid Mech.}{222}{251--265}
 
%S1
\refbook{refS1}
{\author{C.}{Sparrow}}{1982}
            {The Lorenz Equations: Bifurcations, Chaos, and Strange Attractors}
            {Springer}{New York}
 
%S9
\refarticle{refS9}
{\author{E.A.}{Spiegel}}{1987}
            {Chaos: a mixed metaphor for turbulence}
            {Proc. R.~Soc. Lond.}{413A}{87--95}
 
%S20
\refarticle{refS20}
{\author{J.W.}{Swift} \and \author{K.}{Wiesenfeld}}{1984}
            {Suppression of period doubling in symmetric systems}
            {Phys. Rev. Lett.}{52}{705--708}
 
%S30
\refphd{refS30}
{\author{J.}{Swinton}}{1991}
            {Stability of homoclinic waves in lasers and fluids}
            {Imperial College, London}
 
%S6
\refarticle{refS6}
{\author{P.}{Sz\'epfalusy} \and \author{T.}{T\'el}}{1985}
            {Properties of maps related to flows around a saddle point}
            {Physica}{16D}{252--264}
%S100
\refarticle{refS100}
{\author{Emily}{Stone} \and \author{Philip}{Holmes}}{1990}
            {Random perturbations of heteroclinic attractors}
            {SIAM J. Appl. Math.}{50}{726--743}

\refbook{refS101}
{\author{R}{Stratonovich}}{1963}
            {Topics in the theory of random noise}
            {Gordon and Breach}{New York}

%S102
\refarticle{refS102}
{\author{N.G.}{Stocks}, \author{R.}{Mannella} \and \author{P.V.E.}{McClintock}}{1989}
            {Influence of random fluctuations on delayed bifurcations:
        The case of additive white noise}
            {Phys. Rev. A}{40}{5361--5369}

%S103
\refarticle{refS103}
{\author{N.G.}{Stocks}, \author{R.}{Mannella} \and \author{P.V.E.}{McClintock}}{1990}
            {Influence of random fluctuations on delayed bifurcations.
        II. The cases of white and coloured additive 
        and multiplicative noise}
            {Phys. Rev. A}{42}{3356--3362}

\refartbook{refS104}
{\author{R.L.}{Stratonovich}}{1989}
            {Some Markov methods in the theory of stochastic processes
in nonlinear dynamical systems}
            {Noise in nonlinear dynamical systems, Vol. I}
            {\author{Frank}{Moss} \and \author{P.V.E.}{McClintock}}
            {Cambridge University Press}{Cambridge}{16--71}
 
\refartbook{refS105}
{\author{J.M.}{Sancho} \and \author{M.}{San Miguel}}{1989}
            {Langevin equations with coloured noise}
            {Noise in nonlinear dynamical systems, Vol. I}
            {\author{Frank}{Moss} \and \author{P.V.E.}{McClintock}}
            {Cambridge University Press}{Cambridge}{72--109}
 
%S106
\refarticle{refS106}
{\author{N.G.}{Stocks}, \author{N.D.}{Stein} \and \author{P.V.E.}{McClintock}}{1993}
            {Stochastic resonance in monostable systems}
            {J. Phys. A}{26}{L385--L390}

%S107
\refarticle{refS107}
{\author{David}{Sigeti} \and \author{Werner}{Horsthemke}}{1989}
            {Pseudo-regular oscillations induced by external noise}
            {J. Stat. Phys}{50}{1217--1222}

%S108
\refarticle{refS108}
{\author{Steven H.}{Strogatz}, \author{Renato E.}{Mirollo}
\and \author{Paul C.}{Matthews}}{1992}
            {Coupled Nonlinear Oscillators below the Synchronization
Threshold: Relaxation by Generalized Landau Damping}
            {Phys. Rev. Lett.}{68}{2730--2733}

%S109
\refarticle{refS109}
{\author{J.W.}{Swift}, \author{P.C.}{Hohenberg}
 \and \author{Guenter}{Ahlers}}{1991}
            {Stochastic Landau equation with time-dependent drift}
            {Phys. Rev. A}{43}{6572--6580}
 
%S110
\refarticle{refS110}
{\author{John}{Smythe}, \author{Frank}{Moss} \and
\author{P.V.E.}{McClintock}}{1983}
{Observation of a Noise-induced Phase Transition with an Analog Simulator}
            {Phys. Rev. Lett.}{51}{1062--1065}

%S111
\refarticle{refS111}
{\author{V.E.}{Shapiro}}{1993}
            {Systems near a critical point under multiplicative noise 
and the concept of effective potential}
            {Phys. Rev. E}{48}{109--120}

%S112
\refarticle{refS112}
{\author{V.E.}{Shishkova}}{1973}
            {Examination of a system of differential equations with a
small parameter in the highest derivatives}
            {Soviet Math. Dokl.}{14}{183--187}

%S113
\refbook{refS113}
{\author{Joel}{Smoller}}{1983}
            {Shock Waves and Reaction-Diffusion Equations}
            {Springer}{New York}

%S114
\refarticle{refS114}
{\author{D.J.}{Scalapino}, \author{M.}{Sears} \and
\author{R.A.}{Ferrell}}{1972}
        {Statistical Mechanics of One-Dimensional Ginzburg-Landau Fields}
         {Phys. Rev. B}{6}{3409--3416}

%S115
\refarticle{refS115}
{\author{Masuo}{Suzuki}}{1978}
        {Theory of instability, Brownian motion and formation of
macroscopic order}
         {Phys. Lett. A}{67}{339--341}
 
%T4
\refarticle{refT4}
{\author{F.}{Takens}}{1974}
            {Forced oscillations and bifurcations}
            {Comm. Math. Inst., Rijksuniversiteit Utrecht}{3}{1--59}
 
%V1
\refarticle{refV1}
{\author{G.}{Veronis}}{1965}
            {On finite amplitude instability in thermohaline convection}
            {J.~Marine Res.}{23}{1--17}
 
%V4
\refarticle{refV4}
{\author{G.}{Veronis}}{1966}
            {Motions at subcritical values of the Rayleigh number in a rotating
             fluid}
            {J.~Fluid Mech.}{24}{545--554}
 
%V7
\refarticle{refV7}
{\author{G.}{Veronis}}{1966}
            {Large-amplitude B\'enard convection}
            {J.~Fluid Mech.}{26}{49--68}
 
%V4a
\refarticle{refV4a}
{\author{G.}{Veronis}}{1966a}
            {Motions at subcritical values of the Rayleigh number in a rotating
             fluid}
            {J.~Fluid Mech.}{24}{545--554}
 
%V7b
\refarticle{refV7b}
{\author{G.}{Veronis}}{1966b}
            {Large-amplitude B\'enard convection}
            {J.~Fluid Mech.}{26}{49--68}
 
%V5
\refarticle{refV5}
{\author{G.}{Veronis}}{1968}
            {Large-amplitude B\'enard convection in a rotating fluid}
            {J.~Fluid Mech.}{31}{113--139}

%W12
\refarticle{refW12}
{\author{R.W.}{Walden}, \author{P.}{Kolodner},
 \author{A.}{Passner} \and \author{C.M.}{Surko}}{1985}
            {Travelling waves and chaos in convection in binary fluid mixtures}
            {Phys. Rev. Lett.}{55}{496--499}
 
%W16
\refarticle{refW16}
{\author{N.O.}{Weiss}}{1966}
            {The expulsion of magnetic flux by eddies}
            {Proc. Roy. Soc.}{293A}{310--328}
 
%W3
\refarticle{refW3}
{\author{N.O.}{Weiss}}{1981}
            {Convection in an imposed magnetic field. Part 1. The development
             of nonlinear convection}
            {J.~Fluid Mech.}{108}{247--272}
 
%W4
\refarticle{refW4}
{\author{N.O.}{Weiss}}{1981}
            {Convection in an imposed magnetic field. Part~2. The dynamical
             regime}
            {J.~Fluid Mech.}{108}{273--289}
 
%W3a
\refarticle{refW3a}
{\author{N.O.}{Weiss}}{1981a}
            {Convection in an imposed magnetic field. Part~1. The development
             of nonlinear convection}
            {J.~Fluid Mech.}{108}{247--272}
 
%W4b
\refarticle{refW4b}
{\author{N.O.}{Weiss}}{1981b}
            {Convection in an imposed magnetic field. Part~2. The dynamical
             regime}
            {J.~Fluid Mech.}{108}{273--289}
 
%W11
\refarticle{refW11}
{\author{N.O.}{Weiss}, \author{D.P.}{Brownjohn},
 \author{N.E.}{Hurlburt} \and \author{M.R.E.}{Proctor}}{1990}
            {Oscillatory convection in sunspot umbrae}
            {Mon.~Not. R.~Astr. Soc.}{245}{434--452}

%W5
\refbook{refW5}
{\author{S.}{Wiggins}}{1988}
    {Global Bifurcations and Chaos: Analytical Methods}
    {Springer}{New York}
 
%V100
\refarticle{refV100}
{\author{S.Ya.}{Vyshkind} \and \author{M.I.}{Rabinovich}}{1976}
            {The phase stochastization mechanism and the structure of
wave turbulence in dissipative media}
            {Sov. Phys. JETP}{44}{292--299}
%W100
\refarticle{refW100}
{\author{J.-M.}{Wersinger}, \author{J.M.}{Finn}
  \and \author{E.}{Ott}}{1980}
            {Bifurcation and ``strange'' behaviour in instability
saturation by nonlinear three-wave mode coupling}
            {Phys. Fluids}{23}{1142--1154}

%note1
\refnote{refNote1}
{The change of variables from the equations for
the amplitudes of the three wave modes used here differs  from
that of Hughes and Proctor; our $(x,y,z)$ corresponds to $(-x,w,y)$}

%note2
\refnote{refNote2}
{The distribution of a variable whose logarithm is normally distributed is 
sometimes called lognormal.
In an earlier publication[3], we were unaware of
this convention and used the term `lognormal` for the distribution of
the logarithm of $\ha \ln(n^2)$
 where $n$ is a Gaussian random variable with unit variance.}

%note3
\refnote{refNote3}{In an earlier publication~[12] we called this the
``log-normal'' distribution, being unaware that that term is
conventionally used for the distribution of a variable whose logarithm
is normally distributed}

%note4
\refnote{refNote4}{Hughes and Proctor (reference [5]) performed a
similar reduction. The variables $x,y,z$ used here are related to the
$A,B,C$ used by them via $x=A$, $y=\ha(B+C)$ and $z=\ha(B-C)$; $\delta
= \ha(1-\lambda)$. This choice of variables used in this work
simplifies analysis because of the symmetry (6)}

\refarticle{refT100}
{\author{M.C.}{Torrent} \and \author{M.}{San Miguel}}{1988}
           {Stochastic-dynamics characterization of delayed laser
threshold instability with swept control parameter}
           {Phys. Rev. A}{38}{245--251}

\refarticle{refT101}
{\author{M.C.}{Torrent}, \author{F.}{Sagu\'es} \and \author{M.}{San Miguel}}{1989}
           {Dynamics of sweeping through an instability: Passage-time
statistics for colored noise}
           {Phys. Rev. A}{40}{6662--6672}

\refarticle{refT102}
{\author{M.}{Tlidi}, \author{Paul}{Mandel}  \and
\author{R.}{Lefever}}{1994}
           {Localized Structures and Localized Patterns in Optical Bistability}
           {Phys. Rev. Lett}{73}{640--643}

\refbook{refW101}
{\author{David}{Williams}}{1991}
            {Probability with Martingales}
            {Cambridge University Press}{Cambridge}

\refarticle{refV100}
{\author{Jorge}{Vi\~nals}, \author{Hao-Wen}{Xi}
  \and \author{J.D.}{Gunton}}{1992}
            {Numerical study of the influence of forcing terms
and fluctuations near onset on the roll pattern in Rayleigh-B\'enard
convection in a simple fluid}
         {Phys. Rev. A}{46}{918-927}

\refarticle{refW102}
{\author{Jean-Marie}{Wersinger}, \author{John M.}{Finn}
  \and \author{Edward}{Ott}}{1980}
            {Bifurcations and strange behaviour in instability
saturation by nonlinear mode coupling}
         {Phys. Rev. Lett.}{44}{453--457}

\refartbook{refW103}
{\author{J.B.}{Walsh}}{1986}
            {An introduction to stochastic partial differential equations}
            {Ecole d'\'et\'e de probabilit\'es de St-Flour XIV}
            {\author{P.L.}{Hennequin}}
            {Springer}{Berlin}{pp266--439}

\refphd{refW104}
{\author{R.A.}{Weston}}{1989}
            {Lattice Field Theory and Statistical-Mechanical Models}
            {University of Cambridge}
 
%W105
\refarticle{refW105}
{\author{K.}{Wiesenfeld} \and \author{Bruce}{McNamara}}{1985}
            {Period-Doubling Systems as Small-Signal Amplifiers}
            {Phys. Rev. Lett.}{55}{13--15}

%W106
\refartbook{refW106}
{\author{K.}{Wiesenfeld}}{1989}
            {Period doubling bifurcations: what good are they?}
            {Noise in nonlinear dynamical systems, Vol. II}
            {\author{Frank}{Moss} \and \author{P.V.E.}{McClintock}}
            {Cambridge University Press}{Cambridge}{pp145--178}

%W107
\refarticle{refW107}
{\author{K.}{Wiesenfeld}}{1985}
            {Virtual Hopf phenomenon: A new precursor of
period-doubling bifurcations}
            {Phys. Rev. A}{32}{1744--1752}

%W108
\refarticle{refW108}
{\author{K.}{Wiesenfeld} \and \author{N.F.}{Pederson}}{1987}
            {Amplitude calculation near a period-doubling bifurcation:
        An example}
            {Phys. Rev. A}{36}{1440--1444}
%W109
\refarticle{refW109}
{\author{C.}{DeWitt-Morette} \and \author{K.D.}{Elworthy (eds)}}{1987}
            {New stochastic methods in physics}
            {Phys. Rep.}{77}{121--382}

\refbook{refW110}
{\author{L.C.G.}{Rogers} \and \author{David}{Williams}}{1987}
            {Diffusions, Markov Processes and Martingales.
        Vol 2: \ito\ calculus}
            {Wiley}{Chichester}

%Y100
\refarticle{refY100}
{\author{Marc}{Yor}}{1989}
            {On stochastic areas and averages of planar Brownian motion}
            {J. Phys. A}{22}{3049--3057}

%Z100
\refarticle{refZ100}
{\author{H.}{Zeghlache}, \author{Paul}{Mandel} \and \author{C.}{van den Broeck}
}{1989}
           {Influence of noise on delayed bifurcations}
           {Phys. Rev. A}{40}{286--294}

%Z101
\refartbook{refZ101}
{\author{Paul}{Mandel}, \author{H.}{Zeghlache} \and
\author{T.}{Erneux}}{1989}
           {Time-dependent phase transitions}
        {Far from Equilibrium phase Transitions}
{\author{Luis}{Garrido}}{Springer}{Berlin}{pp218--236}

\endgroup
\inrefsfalse

\spaceandahalf

\damtpaddress

%\doublespace
%\figuresatendfalse
%\figuresatend
%\nofigs

\windowcover

\def\theauthor{G.D.~Lythe}

\title{Dynamics controlled by additive noise}

\signature

\abstract{
Analysis is presented of a system whose dynamics are dramatically
simplified by tiny amounts of additive noise.
 The dynamics divide naturally into two phases.
In the slower phase, trajectories are close to an
invariant manifold; this allows small random disturbances
 to exert a controlling influence.  
 A map is derived which provides an accurate description of the trajectories.}

\publishedin{Nuovo Cimento D {\bf 17} 855--861 (1995)}

%\toappearin{Nuovo Cimento D}

\begingroup
\ShowTP
\endgroup

\pageno=1

\beginsect Introduction

In certain systems described by differential equations, tiny amounts
of added noise have a controlling influence on the 
dynamics.
%\addfourrefs{refH101}{H101}{refH102}{H102}{refM104}{M104}{refS100}{S100}
The influence of noise is a simplifying one: the larger the
noise level the more regular trajectories become. 
Sensitivity 
to noise arises because solutions of the differential
equations spend most of their time near a slow invariant manifold.
 Noise with magnitude $\ep$
 controls the dynamics if $\ep > e^{{-1 \over \mu}}$, where $1/\mu$
is the timescale for dynamics near the invariant manifold.

The type of behaviour described here is found, for example, in
differential equations modelling the resonant interaction of wave
modes\addref{refL101}{L101}, turbulent flows\addref{refS100}{S100},
pulsating laser
oscillations\addref{refM104}{M104} and plane Poiseuille
flow\addref{refF101}{F101}. In this work, the following system,
derived as a model for the shear instability of tall convection
cells\addref{refH102}{H102}, is analysed:
$$\eqalign{ \dot x &= \mu x - y^2+z^2,\cr
 	    \dot y &= y(x-1+\delta) + \delta z,\cr
 	    \dot z &= z(-x-1+\delta) + \delta y;\cr
	    0<&\mu<1,\qquad0\le\delta<0.5.}
 \eqno \myeqno \myeqlabel{mysic}$$

  The dynamics in the case $\delta=0$
were analysed by Hughes
and Proctor in terms of a \od\ map\addref{refH102}{H102}.
In this work a map is derived which describes the solutions
when $\delta$ is non-zero and the system is
subject to white noise with magnitude $\ep$ such that
$\ep\ll\sqrt{\mu}\ll 1$. The dynamics are organised about the
 line $y=z=0$ (the unstable manifold of the fixed point at the origin) and,
 to lowest order in $\mu$ and $\delta^2$, can be
described exclusively in terms of the parameter $\mle$.

\figure 6truein  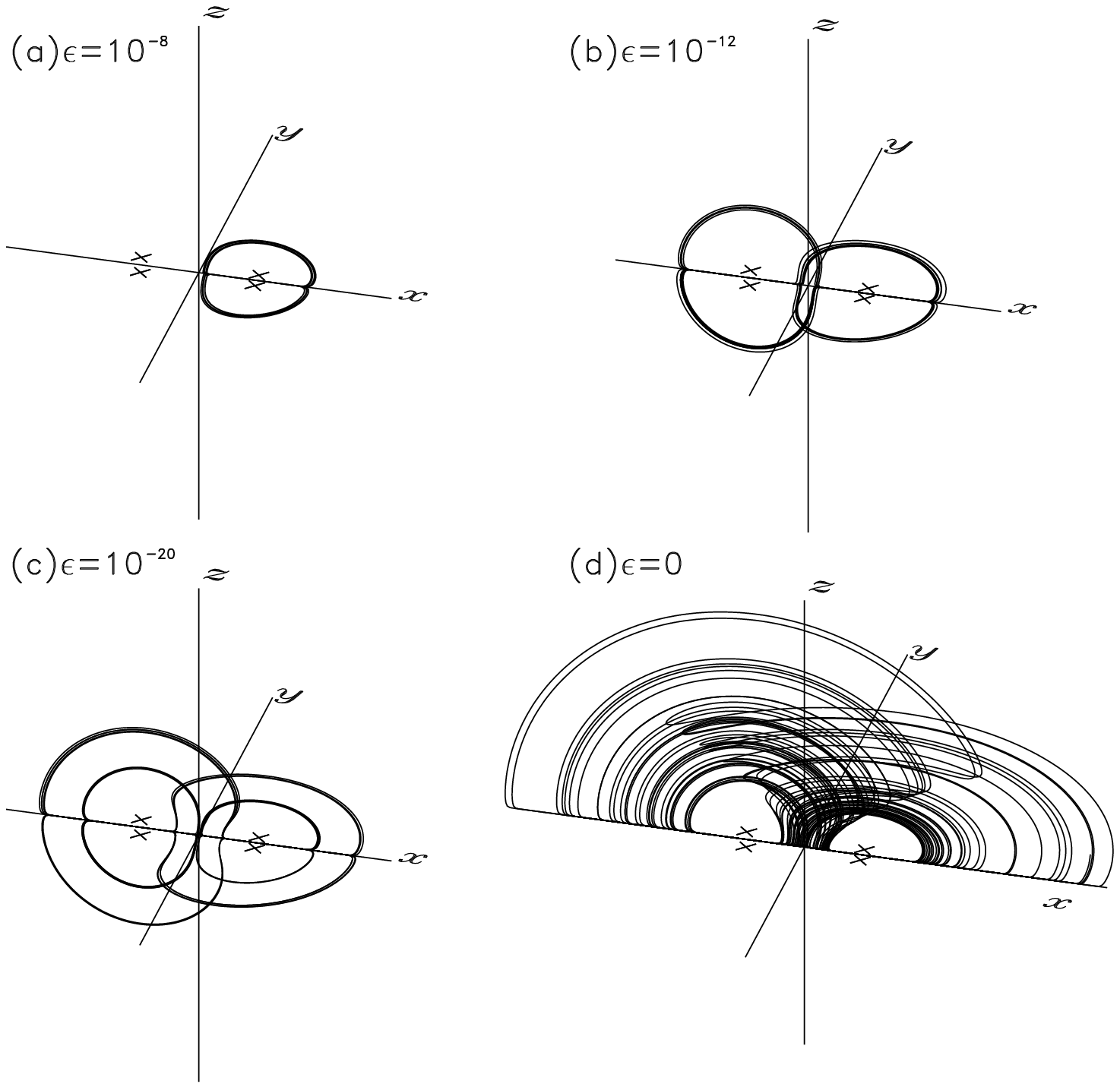
{{\it Noise-dependent solutions.}
The four trajectories shown are solutions of \mysic, differing only in
the level of added noise, $\ep$. The greater the noise level, the simpler the
solution. Sensitivity to noise arises because the solutions spend
most of their time near $y=z=0$ with $|x|$ slowly increasing.
 ($\delta=0.3$ and $\mu=0.01$.) Fixed points are marked with crosses.}
\myfigurelabel{fdzt}

In the simplest case, Figure 1(a), the solutions consist of long
periods (slow phases) during which $y$ and $z$ are $\Or{\ep}$ and $|x|$
slowly increases, occasionally interrupted by fast phases during which
$|x|$ decreases. The solution is then fully specified in terms of the \pd\
of $\xm$, the turning point of $x$ at the end of a slow phase. An accurate
expression for this distribution
is obtained by modelling the slow phase with
the stochastic differential equation
$$dy = yg(x)dt + \ep dw
 \eqno \myeqno \myeqlabel{ysde}$$
where $w$ is the Wiener process\addref{refG100}{G100} and 
 $x$ is taken to be a function of time satisfying $\dot x =\mu x$.
The function $g(x)$, which approximates $x(t)-1+\delta+{z\over y}(t)$,
begins negative and passes through $0$ in the course of a slow phase.
The ratio $r(t)={z\over y}$ is obtained by assuming that it is
slaved to the instantaneous value of $x$.
Because $g(x)$ is slowly-varying, \ysde\ is a dynamic bifurcation problem
in which trajectories remain near $y=z=0$ for a
 time $\Or{{1\over \sqrt{\mu}}}$ after $g(x)$ becomes
positive. Note that it is only by the action of noise that $y$ changes sign.
The condition $\ep\ll\sqrt{\mu}$ is necessary to justify the
neglect, for long periods, of the quadratic terms in the equation for
$\dot x$. The analysis, based on separating slow and fast phases,
is accurate when $\mu\ll 1$.

The solution of \ysde\ specifies the mean and standard deviation
of $y$ as a function of $x$. It is thus possible
to calculate, as a function of $x$,
 the probability that $y^2-z^2>\mu x$; taking the
derivative with respect to $x$ of this probability gives the distribution
of $\xm$, which has a non-Gaussian form\addref{refL100}{L100}.
The maximum of the distribution is at $\xm=\hat x$ where,
to lowest order in $\mu$,  $\hat x$ satisfies
$$F(\hat x) -F(\alpha)= \mle
\eqno \myeqno \myeqlabel{xhatlow}$$
where
${d \over dx}F(x)=g(x)/x$ and $g(\alpha)=0$.
 The standard deviation of the distribution of $\xm$ is
$$ \sigma_{\xm}={\pi \over 2\sqrt{2}}{\mu \hat x \over g(\hat x)} .
\eqno \myeqno \myeqlabel{sdxm}$$
Note that the width of the distribution is $\Or{\mu}$, not $\Or{\ep}$.
The \pd\ of $\xm$ is affected only slightly by the
introduction of
a finite correlation time (`colour') in the noise\addref{refL108}{L108}.
 On the other hand, multiplicative noise or mean-zero
deterministic perturbations have a much less dramatic effect 
on trajectories than additive Gaussian noise.

For the trajectory shown in Figure 1(a),
successive values of $\xm$ are independent and
the specification of the \pd\ of $\xm$ provides an essentially
complete description of the solution of \mysic.  More
generally the value $\hat x$ is an upper limit 
on $\xm$. This still has a profound, and simplifying, influence on the
dynamics. 

A different approximation scheme is appropriate for the fast phase.
The simplest is to assume that the brief excursions away from
the vicinity of $y=z=0$ are semicircular, ie
that $x_0$ is related to the previous value of $\xm$ by
$$\xm - 1 +\delta = 1-\delta - x_0,
\eqno \myeqno \myeqlabel{simplex}$$
where $x_0$ is the minimum value of $x$ in a cycle
(and thus the starting point for a slow phase).
 Corrections to \simplex, necessary for an accurate
description of the dynamics in the cases shown in Figure 1(c) and 1(d),
will be introduced in the next section. They are not necessary in the case
shown in Figure 1(a) 
because, in every slow phase, $|y|$ and $|z|$ 
descend to $\Or{\ep}$ before $g(x)$ passes through $0$ and so
$\xm$ is in fact independent of the previous $x_0$.
The trajectory shown in Figure 1(b) differs from 1(a)
in that $x$ changes sign in the course of each fast
phase. Because of the invariance of \mysic\ under the transformation
$$x\mapsto-x,\qquad{\rm and}\qquad y \leftrightarrow z,
 \eqno \myeqno \myeqlabel{trans}$$
the analysis of each slow phase can proceed for $x<0$ as before, with
the roles of $y$ and $z$ reversed and $x$ replaced by $|x|$.
Thus the statistics of $|\xm|$ are the same as in Figure 1(a),
and successive values of $\xm$ have opposite signs.
(At the same parameter values as Figure 1(a), there is a stable noisily
periodic orbit with $x$ always negative.)

The change from trajectories in which $x$ always keeps the same sign
(Figure 1(a)) to those in which $x$ changes sign (Figure 1(b))
occurs when $\hat x$ is sufficiently large, 
ie when $\mle$ is sufficiently large.
The next qualitative change that occurs for larger $\mle$
(larger $\mu$ or smaller $\ep$)
 is a period-doubling bifurcation (Figure 1(c)). This occurs when 
the value of $|x_0|$ that follows $\hat x$
approaches $1-\delta$ so that $y$ and $z$ can
remain above the noise level throughout a slow phase. 
The condition for this bifurcation also translates into a condition on
$\mle$.  After the period-doubling bifurcation, 
the parameter $\mle$ can still be
used as a bifurcation parameter but trajectories
can no longer simply be described in terms of one probability 
distribution. 
 A more sophisticated analysis, in terms of a
map, is appropriate. This is developed in the next section.

\beginsect Description by an analytical map

Starting from a given value of $\xm$, there is a fast phase which
carries the system to a value of $x$, $x_0$, at which a slow phase
begins. This slow phase carries the system to a new value of $\xm$.
The object is to describe analytically solutions of \mysic\ in terms
of a one-dimensional map of successive values of $\xm$ (values of $x$
at the end of a slow phase at which $\dot x=0$). The strategy is to
treat the slow phase, in which noise may have a controlling
influence, separately from
the fast phase, in which noise is unimportant but the
nonlinearities are more troublesome.

Analysis of the fast phase is based on the function $h(t)$ where
$h=(x-1+\delta)^2 +y^2-z^2$.
This satisfies
$\dot h = 2\mu(x-1+\delta)x +4xz^2$.
 (Typically $z < \delta$ during the fast phase.)
By integrating $\dot h$ along a semicircular
trajectory, the following approximation was obtained:
$$(|x'|-1+\delta)^2 = (|\xm|-1+\delta)^2  + (|\xm|-1+\delta)(6\mu+4\delta^2).
 \eqno \myeqno \myeqlabel{fastform}$$
Here $x'$ is written instead of $x_0$ because, if $x'$ and $\xm$ have
opposite signs,  a new slow phase does not begin immediately --
notice the looping back in Figure 1(d).
Given $\xm$,
to generate $x_0$ (the starting value for the next slow phase)
further iterates of \fastform, $x'', x''',\ldots$ are generated.
 (These are approximations for
successive turning points of $x$.)
When two successive values have the same sign, the procedure ends and
the system is deemed to have reached $x_0$. Analysis of the fast
phase is simpler when $\delta$ is $0$ or is extremely
small\addref{refH102}{H102} because looping back of trajectories
does not occur.

The slow phase starts with $x=x_0$.
If $|x_0|$ is sufficiently far from
 $1-\delta$ that $|y|$ and $|z|$ descend
to the noise level before $g(t)$ passes through $0$, the \pd\ of $\xm$
can be described very accurately by \xhatlow\ and \sdxm. On the other hand, if
the minimum value of $|y|$ in a slow phase
 is above the noise level then the
relationship between $\xm$ and the previous value of $x_0$ is 
$$F(x_0)=F(\xm).
 \eqno \myeqno \myeqlabel{detxm}$$

It is possible to derive an expression which has both \xhatlow\ and
\detxm\ as limits\addref{refL100}{L100}. 
The changeover between \regime s 
is very rapid as a function of $x_0$, occurring when
the minimum value of $y$ that would be found in a slow phase
without noise (typically $e^{-1/\mu}$) equals the noise level $\ep$.
In practice, therefore, $\xm$ is generated by taking the minimum
of the two values given by \xhatlow\ and \detxm\
-- the effect of noise, if any, is always to reduce the average value
of $\xm$.
The map thus obtained is compared with
that resulting from numerical solution of \mysic\ in Figure 2, where
a rather low noise level is chosen to exhibit some of the structure of
the map. For larger values of $\ep$, the cutoff value of $\xm$,
given by \xhatlow, means that only a few humps are seen.

\figure 4.5truein  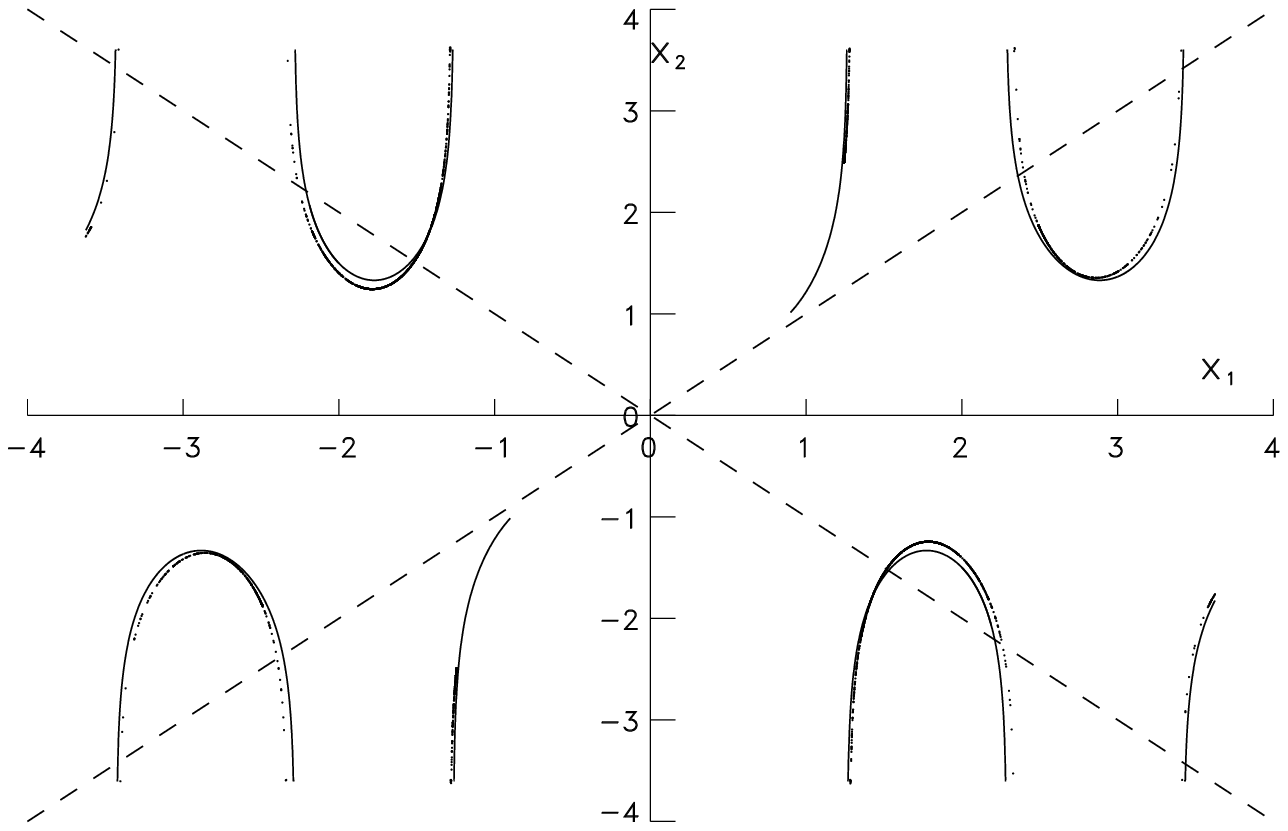
{{\it Map of successive values of $\xm$.} If $x_1$ is the turning
point of $x$ at the end of a slow phase then the next such turning
point is $x_2$.
The dots are numerical results from simulation of \mysic, with
$\ep=10^{-80}$ noise added independently to each variable,
and the solid lines calculated as
described in the text.  ($\mu=0.01$, $\delta=0.3$.)
The dotted line is $x_1=x_2$.
The most important effect of noise is to set an upper limit,
(in this case 3.6) on $\xm$.
 This reduces the map to a finite number of humps.
(In numerical simulations of trajectories,
 a value of $x$ at which $\dot x=0$ is deemed to be at the
end of a slow phase if, in addition, $\dot y/y>0$ and $\dot z/z>0$.
This has the effect of restricting the maps to $|\xm| > \sqrt{1-2\delta}$.)
}
\myfigurelabel{fdzt}

The bifurcation structure of multi-modal maps is extremely rich, but
the smearing in $x$ due to \sdxm\ means that, 
for non-vanishing $\mu$ only the first few bifurcations are distinguished.
 Finite $\mu$ also has the effect of making the map
inaccurate in the region $||\xm|-1+\delta|<\sqrt{\mu}$,
where the distinction between fast and slow phases breaks down.

The numerical solutions exhibited in this work were produced using
a second order stochastic algorithm\addref{refK100}{K100}.
Adding very small noise usually presents no difficulties
because the noise-sensitive variables are themselves small, but for
extremely small noise levels, it may be necessary to use $\ln y$ and
$\ln z$ as variables rather than $y$ and $z$.

\beginsect Relevance to shear in convection

When a horizontal layer of fluid is heated from below, convection
begins if the temperature difference across the layer exceeds a
threshold value. Just above this threshold, the motion often consists
of rolls\addref{refC106}{C106}. If these rolls are tall and thin they
can exist for long periods but
are typically subject to shear (a net sideways movement)
 which can temporarily destroy the roll pattern.
 In this situation, the motion of the fluid 
is in two space dimensions and can therefore be described in terms of
a scalar field known as the stream function. The stream function
in this case can
be written as a sum of three important modes (representing
`convection', `shear' and `tilt') plus other slaved 
modes\addref{refM110}{M110}. Substitution of this form for the stream
function into the
partial differential equations for Boussinesq convection results in a
third order system of ordinary differential equations which
successfully models the behaviour of the solutions of 
the partial differential equations when the rolls
are tall and the Prandtl number is small.
The system \mysic\ is produced from this by rescaling 
the three variables so that
the coefficients of all quadratic terms are unity\addref{refNote4}{Note4}.

In the model system \mysic, the variable $x$ is proportional to the
amplitude of the convective roll pattern, which persists
for long periods and is occasionally destabilised by shear before
reestablishing itself. The variables $y$ and $z$ are linear
combinations of `shear' and `tilt', $\mu$ is proportional to the amount
by which the temperature difference across the layer exceeds the
critical value, and $\delta$ is a function of the Prandtl number.
(A fifth order set of ordinary differential equations is appropriate
when the Prandtl number is not small and
there is a magnetic field present\addref{refM110}{M110}.)

\beginsect Conclusion

Dynamical systems whose solutions return repeatedly to the
neighbourhood of a slow invariant manifold can be controlled by added noise
with magnitude $\ep$ if $\ep<e^{-1/\mu}$, where ${1\over\mu}$ is the 
timescale for dynamics near the manifold.
It is possible to construct analytically a one-dimensional map
when $\ep\ll\sqrt{\mu}\ll 1$. Noise affects the dynamics by setting
 an upper limit on the size of excursions away from the
invariant manifold. The upper limit is a function of $\mle$ and
the width of the appropriate \pd\ is
proportional to $\mu$ (not to $\ep$). 

\beginack
\noindent I have benefited from discussions with 
Paul Matthews, Alastair Rucklidge and Mike Proctor.

\smallskip

\showrefs

\Trailers

\end